\documentclass[11pt,a4paper]{article}
\usepackage{jheppub}
\usepackage{amsmath}
\usepackage{amssymb}

\title{Systematics of Higher-spin Light-front Interactions}
\author[a,1]{Anders K. H. Bengtsson\note{Work supported by the Research and Education Board at the University of Bor{\aa}s.}}

\affiliation[a]{School of Engineering, University of Bor{\aa}s, All\' egatan 1, SE-50190 Bor\aa s, Sweden.}

\emailAdd{anders.bengtsson@hb.se} 

\abstract{The original cubic interaction terms for higher spin gauge fields in four dimensions and their reformulation using Fock space vertex operators is reviewed. As a new result, the complete list of all cubic vertex functions in $D=4$ is derived. It is observed, contrary to what would have been expected, that the non-linear dynamical Poincar{\'e} transformations do not restrict the cubic interactions beyond what is required by kinematics. The role of the $\mathrm{SU}(1,1)$ algebra of tracelessness constraints is clarified. It is shown that higher spin fields couple non-minimally to gravity at the cubic level in $D=4$ light-front dynamics. Based on a detailed analysis of the structure of the light-front Poincar{\'e} algebra, the formalism is then generalized to all orders in the interaction. The interacting theory, being a deformation of the free theory, takes the form of a $L_\infty$ algebra. It is conjectured that the ensuing recursive equations, if they can be explicitly solved, will encode all interaction data into a denumerable set of functions of $p^+$, the overall transverse momentum structure being fixed already at the kinematical level.}


\begin{document}

\maketitle
\pagebreak

\section{Introduction}\label{sec:Introduction}
A characteristic feature of the light-front cubic interactions for massless higher helicity fields is the simple binomial expansion form they take \cite{BBB1983a}
\begin{equation}\label{eq:CubicInteractionLightFront}
\int d^4x\sum_{n=0}^{\lambda}(-1)^n{\lambda\choose n}(\partial^+)^\lambda\phi\Big[{\partial\over\partial^+}\Big]^{(\lambda-n)}{\bar\phi}\Big[{\partial\over\partial^+}\Big]^n{\bar\phi}+\mbox{c.c.}
\end{equation} 
This expression, with gauge group structure constants (anti-symmetrization) understood for odd helicity, appears somewhat mysteriously out of the deformed Poincar{\'e} algebra. It becomes more clear when the interaction terms are reformulated in momentum Fock space in terms of vertex operators \cite{BBL1987}. In that formulation, the momentum structure for the helicity $\lambda$ cubic interaction is essentially given by $\mathbb{P}^\lambda$ where $\mathbb{P}$ is defined by
\begin{equation}\label{eq:DefBlackboardMomentum}
\mathbb{P}=-\frac{1}{3}\sum_{r=1}^3\widetilde\gamma_r p_r\quad\text{with}\quad \widetilde\gamma_r=\gamma_{r+1}-\gamma_{r+2}\quad\text{and}\quad\gamma=p^+
\end{equation}
This result seems not to be very well known.\footnote{An exception is work by Metsaev, see for instance \cite{Metsaev2005ar}.} The purpose of the present paper is to explicate on the 1987 work and advance the treatment. The structure of the cubic vertices will be clarified and the full solution to that order in the Poincar{\'e} algebra will be derived (see formulas \eqref{eq:VertexFunctionsGeneral}). It will be observed that the kinematics of the vertex suffice to determine it. Cubic couplings involving different higher helicities will be exhibited, as an example non-minimal spin $4$ coupling to gravity (spin $2$).\footnote{Such couplings were implicit in the 1987 paper \cite{BBL1987}.}

Light-front higher helicity dynamics in $D=4$ still lack systematics, and hopefully the present work will contribute towards amending that deficiency. The main purpose is to prepare the way for a computation of quartic interactions. For that reason the paper is quite detailed.\footnote{Also, light-front calculations involve some calculational subtleties, in particular in momentum space, that are easily over-looked on a first attempt.} I've tried to design notation to enhance readability and computability, which however, is not so easy given the inherent complexity of higher spin theory.

The present text is organized as follows. Section \ref{sec:Rationale} presents the rationale for working in momentum space. In section \ref{sec:LFHigherHelicityFields}, the Fock space of higher helicity fields is set up. Section \ref{sec:PoincareAlgebra} contains a systematic discussion of the $D=4$ light-front Poincar{\'e} algebra. Section \ref{sec:ReviewCubicInteraction} reviews the cubic vertex with some examples, for instance non-minimal couplings of higher spin to gravity. Sections \ref{sec:Framework} and \ref{sec:ComputationDDDifferential} constitute the main body of the paper where a framework capable of supporting a computation of quartic and higher order vertices is set up. In section \ref{sec:SystematicsCubicVertices}, the complete cubic vertex is derived. The last section, \ref{sec:ConventionsFormulas}, summarizes the light-front conventions as well as some elaborations of material from the text.

\section{Rationale for momentum space}\label{sec:Rationale}
Let us begin by transforming \eqref{eq:CubicInteractionLightFront} from configuration space to momentum space using
\begin{equation}\label{eq:MomentumFieldFourierTransformFull}
\phi(x^+,x^-,x,\bar{x})=\frac{1}{(2\pi)^{3/2}}\int dp^+dpd\bar{p}\phi(x^+,p^+,p,\bar{p})e^{i(-p^+x^-+p\bar{x}+\bar{p}x)}
\end{equation}
or in shorthand
\begin{equation}\label{eq:MomentumFieldFourierTransformShort}
\phi(x)=\frac{1}{(2\pi)^{3/2}}\int d^3p\phi(p)e^{ip\cdot x}
\end{equation}
where we understand that both $x$\,-\,space and $p$\,-\,space fields depend on the light-front time $x^+$. Evolution in $x^+$ is generated by the light-front Hamiltonian $h=p^-$. Now inserting \eqref{eq:MomentumFieldFourierTransformShort} into \eqref{eq:CubicInteractionLightFront} yields
\begin{equation}\label{eq:CubicInteractionMomentumTransformed}
\begin{split}
&\frac{1}{(2\pi)^{9/6}}\int dx^+d^3x\int\prod_{r=1}^3 d^3p_r e^{ip_r\cdot x}\times\\
&\sum_{n=0}^{\lambda}(-1)^n{\lambda\choose n}(ip_1^+)^\lambda\phi(p_1)\Big[{ip_2\over ip_2^+}\Big]^{(\lambda-n)}{\bar\phi(p_2)}\Big[{ip_3\over\ ip_3^+}\Big]^n{\bar\phi(p_3)}+\mbox{c.c.}\\
&=\frac{1}{(2\pi)^{3/2}}\int dx^+\prod_{r=1}^3 d^3p_r\delta^3\Big(\sum_{t=1}^3 p_t\Big)\times\\
&i^{2\lambda}\sum_{n=0}^{\lambda}{\lambda\choose n}(p_1^+)^\lambda\Big[{p_2\over p_2^+}\Big]^{(\lambda-n)}\Big[-{p_3\over\ p_3^+}\Big]^n\phi(p_1){\bar\phi(p_2)}{\bar\phi(p_3)}+\mbox{c.c.}\\
&=\frac{(-1)^\lambda}{(2\pi)^{3/2}}\int dx^+\prod_{r=1}^3 d^3p_r\delta^3\Big(\sum_{t=1}^3 p_t\Big)(p_1^+)^\lambda\Big[{p_2\over p_2^+}-{p_3\over\ p_3^+}\Big]^\lambda\phi(p_1){\bar\phi(p_2)}{\bar\phi(p_3)}+\mbox{c.c.}
\end{split}
\end{equation}
Using cyclic symmetry in field labels $1,2,3$ and momentum conservation, we get for the momentum structure
\begin{equation}\label{eq:CubicMomentumStructure}
(-1)^\lambda(p_1^+)^\lambda\Big[{p_2\over p_2^+}-{p_3\over\ p_3^+}\Big]^\lambda=\Big(\frac{\gamma_1}{\gamma_2\gamma_3}\mathbb{P}\Big)^\lambda
\end{equation}
which is the precise form of what we claimed above. For instance, the Yang-Mills cubic momentum vertex comes out directly for $\lambda=1$. The gravity cubic vertex comes out as the square of the Yang-Mills cubic vertex. The momentum factor in \eqref{eq:CubicMomentumStructure} can be written in terms of the bracket notation of MHV-amplitudes since using cyclic symmetry in field labels we have the relation $\mathbb{P}=-\sqrt{\gamma_1\gamma_2}\langle p_1\,p_2\rangle/\sqrt2$ (see section \ref{subsec:AmplitudeNotation} for details).

In trying to answer the question of how this generalizes to higher orders, the momentum space representation is in my opinion more convenient than working in configuration space with derivatives on fields. The fields can then essentially be dropped from consideration since all interaction data must be encoded in functions of the momenta. The structure, to the extent that it can be presently unraveled, is still complicated though. If the light-front approach is at all feasible, we still need further organizing principles for higher than cubic order.

\section{Light-front higher helicity fields}\label{sec:LFHigherHelicityFields}
We work in four space-time dimensions to simplify as much as possible. In this dimension, all gauge fields have $2$ physical degrees of freedom. This simplifies the mathematics since a complexified notation can be used throughout. Thus we work with the field $\phi_\lambda$ and its complex conjugate $\bar\phi_\lambda$ of helicities $\lambda$ and $-\lambda$ respectively. The wave equations are simply
\begin{equation}
\begin{split}
\partial^-\phi_\lambda&=\frac{\partial\bar\partial}{\partial^+}\phi_\lambda\\
\partial^-\bar\phi_\lambda&=\frac{\partial\bar\partial}{\partial^+}\bar\phi_\lambda
\end{split}
\end{equation}
irrespective of helicity. Indeed, gauge fixing the four-dimensional Fronsdal \cite{Fronsdal1978} equations for arbitrary spin, yield these equations. Helicity data is maintained by the Lorentz generators in transverse directions. 

We introduce a two-dimensional complex internal Fock space spanned by oscillators $\alpha^\dagger$ and $\bar\alpha^\dagger$ where
\begin{equation}\label{eq:OscillatorCommutators}
[\alpha,\bar\alpha^\dagger]=[\bar\alpha,\alpha^\dagger]=1
\end{equation}
Using this we collect all helicities in a Fock space field
\begin{equation}\label{eq:FockSpaceField}
\vert\Phi(p)\rangle=\sum_{\lambda=0}^\infty\frac{1}{\sqrt{\lambda!}}\left(\phi_\lambda(p)(\bar\alpha^\dagger)^\lambda+\bar{\phi}_\lambda(p)(\alpha^\dagger)^\lambda\right)\vert 0\rangle
\end{equation}
In formulas like this, $p$ is short for $p,\bar{p}$ and $\gamma=p^+$. This Fock space field is real in the sense that
\begin{equation}\label{eq:RealityFockField}
\begin{split}
\vert\Phi(p)\rangle^\dagger&=\langle\Phi(p)\vert\\
&=\sum_{\lambda=0}^\infty\frac{1}{\sqrt{\lambda!}}\langle 0\vert\left(\bar{\phi}_\lambda(p)(\alpha)^\lambda+\phi_\lambda(p)(\bar{\alpha})^\lambda\right)
\end{split}
\end{equation}
The fields $\phi$ and $\bar{\phi}$ being functions of $x^+,p^+,p,\bar{p}$, carry mass dimension $-2$. This follows from the Fourier transform \eqref{eq:MomentumFieldFourierTransformFull} since $\phi(x)$ has dimension $1$ as usual. The vacuum and the oscillators are dimensionless, hence the Fock space field $\vert\Phi(p)\rangle$ also carry dimension $-2$. When studying interactions a shorthand notation is practical; $\vert\Phi_r\rangle$ is shorthand for $\vert\Phi(\gamma_r,p_r,\bar{p}_r)\rangle$ expanded over $\alpha_r^\dagger$ and $\bar{\alpha}_r^\dagger$ and correspondingly for $\langle\Phi_r\vert$. In an $\nu$-order interaction term, $r$ will run from $1$ to $\nu$ and serve as a label on the fields.

\subsection*{Tracelessness and SU(1,1)}
The well-known tracelessness requirements from Fronsdal higher spin theory are imposed on the Fock field. In principle there could be mixed excitations such as $\alpha^\dagger\bar\alpha^\dagger$. These are removed by imposing the condition
\begin{equation}\label{eq:TracelessnessCondition}
T|\Phi\rangle=0\quad\text{ with }\quad T=\bar\alpha\alpha
\end{equation}
where $T$ is what remains in the light-front of the tracelessness operator from the covariant BRST approach to $D=4$ higher spin theory \cite{AKHB1988}. Without this restriction the Fock space of the oscillator pair $(\alpha,{\bar\alpha}^\dagger)$ and $(\bar\alpha,\alpha^\dagger)$ is much larger. The full set of fields on excitation level $\lambda$ is
\begin{equation}\label{eq:VerticalFields}
\vert\Omega^{\lambda}\rangle=\sum_{k=0}^\lambda\phi_{\lambda-k,k}({\bar\alpha}^\dagger)^{\lambda-k}(\alpha^\dagger)^{k}\vert0\rangle
\end{equation}
where the physical states of helicity $\lambda$ and $-\lambda$ correspond to $k=0$ and $k=\lambda$ respectively. Fields of the form \eqref{eq:VerticalFields} with a given $\lambda$ will be called {\it vertical} for reasons explained in section \ref{subsec:FockSpace}. The full set of Fock space fields is given by
\begin{equation}\label{eq:FullSetFields}
\vert\Theta\rangle=\sum_{\lambda=0}^\infty\vert\Omega^{\lambda}\rangle
\end{equation}
Another parameterization of the Fock space is in terms of {\it horizontal} fields
\begin{align}\label{eq:HorizontalFields}
\vert\Upsilon^{\lambda}\rangle=\sum_{k=0}^\infty\phi_{k,\lambda+k}({\bar\alpha}^\dagger)^{\lambda+k}(\alpha^\dagger)^{k}\vert0\rangle\\
\vert\bar\Upsilon^{\lambda}\rangle=\sum_{k=0}^\infty\phi_{\lambda+k,k}({\bar\alpha}^\dagger)^{k}(\alpha^\dagger)^{\lambda+k}\vert0\rangle
\end{align}

The trace operator $T$ together with its conjugate $T^\dagger$ and the number operator $N=\alpha^\dagger\bar\alpha+{\bar\alpha}^\dagger\alpha+1$ span an $\mathrm{SU}(1,1)$ algebra commuting with $M$ and therefore with all the Lorentz generators. This algebra act irreducibly on either of the horizontal fields.

It turns out that the interaction vertex produce states of the form $\Omega$ in the form $\Omega_3=\langle\Phi_1\vert\langle\Phi_2\vert V_{123}\rangle$. A projection operator onto physical states can be constructed \cite{BBL1987}.

\subsection*{The Fock space and SO(3,2)}
The pair $(\alpha,{\bar\alpha}^\dagger)$ and $(\bar\alpha,\alpha^\dagger)$ span an oscillator representation of the $\mathrm{SO}(3,2)$ algebra \cite{GunaydinSacliogu1982a}. Its ten generators split into three components $g^{-1}\oplus g^{0}\oplus g^{+1}$ where $g^{0}$ is a $\mathrm{U}(2)$ spanned by 
\begin{equation}\label{eq:G0Algebra}
g^{0}=\begin{cases}M_+=\alpha^\dagger\alpha\\ \bar M_-={\bar\alpha}^\dagger\bar\alpha\\ M=\alpha^\dagger\bar\alpha-{\bar\alpha}^\dagger\alpha\\ N=\alpha^\dagger\bar\alpha+{\bar\alpha}^\dagger\alpha+1\end{cases}
\end{equation}
The other components are spanned according to
\begin{align}\label{eq:G+-Algebras}
g^{-1}&=\begin{cases}T_-=\alpha\alpha\\ \bar T_-=\bar\alpha\bar\alpha\\ T=\bar\alpha\alpha\end{cases}\\
g^{+1}&=\begin{cases}T_+=\alpha^\dagger\alpha^\dagger\\ \bar T_+={\bar\alpha}^\dagger{\bar\alpha}^\dagger\\ T^\dagger={\bar\alpha}^\dagger\alpha^\dagger\end{cases}
\end{align}
The $g_0=\mathrm{U}(2)$ sub-algebra is represented irreducibly on vertical fields of the form \eqref{eq:VerticalFields} in the usual angular momentum way. The basis states can be represented as $\vert j,m_j\rangle$ with $j$ integer or half-integer. The full $\mathrm{SO}(3,2)$ is represented on the Fock space $\vert\Theta\rangle$ of \eqref{eq:FullSetFields}. More details are supplied in section \ref{subsec:FockSpace}. Taking positive powers of the $\mathrm{SO}(3,2)$ generators, the higher spin algebra $\mathrm{hso}(3,2)$ can be built (modulo technical details) \cite{Vasiliev2003ev,Vasiliev2004cm}.

\section{The Poincar{\'e} algebra}\label{sec:PoincareAlgebra}
\subsection{Linear realization}\label{subsec:LinearRealization}
In light-front dynamics, the Poincar{\'e} generators split into a set $\mathcal{K}$ of kinematic generators and a set $\mathcal{D}$ of dynamic generators according to
\begin{align}
\mathcal{K}&=\{\gamma,p,\bar{p}\}\cup\{j,j^{+-},j^+,\bar{j}^+\}\\
\mathcal{D}&=\{h\}\cup\{j^-,\bar{j}^-\}
\end{align}
In a realization as operators acting on free Fock fields, the Lorentz generators can be thought of as a sum $j=l+s$ of an orbital part $l$ and a helicity part $s$. In complex notation, such a linear realization can be expressed as
\begin{equation}\label{eq:FreeLorentzGenerators}
\begin{split}
j&=x\bar{p}-\bar{x}p-iM \hspace{46pt} j^{+-}=i\gamma\frac{\partial}{\partial\gamma}+i\\
j^+&=x\gamma \hspace{106pt} \bar{j}^+=\bar{x}\gamma\\
j^-&=xh+ip\frac{\partial}{\partial\gamma}-\frac{i}{\gamma}Mp \hspace{29pt} \bar{j}^-=\bar{x}h+i\bar{p}\frac{\partial}{\partial\gamma}+\frac{i}{\gamma}M\bar{p}\\
\end{split}
\end{equation}
where the Hamiltonian is $h=p{\bar p}/\gamma$ and the helicity contributions to $j$, $j^-$ and $\bar{j}^-$ is carried by
\begin{equation}\label{eq:DefM}
M=\alpha^\dagger\bar{\alpha}-\bar{\alpha}^\dagger\alpha
\end{equation}
This is a hybrid notation where we think of $x$ and $\bar{p}$ as first quantized with $[x,\bar{p}]=[\bar{x},p]=i$ but we treat $x^-$ and $p^+$ explicitly in terms of 
\begin{equation}\label{eq:DefBetaDBeta}
p^+=\gamma\quad\quad\text{ and }\quad\quad x^-=-i\frac{\partial}{\partial\gamma}
\end{equation}
This turns out to be convenient \cite{NLinden1986}\footnote{We differ from \cite{BBL1987} which used $\beta=2p^+$.}. The Hermiticity properties of the Lorentz generators are commented on in section \ref{subsec:HermiticityProperties}.

The precise connection between these generators and field operators can be worked out as follows. The basic equal time light-front field commutator is \cite{NevilleRohrlich1971}
\begin{equation}\label{eq:BasicTimeCommutator}
[\phi(x),\partial_y^+\bar{\phi}(y)]_{x^+=y^+}=i\delta^3(x-y)
\end{equation}
It translates into the corresponding momentum space equal time commutator
\begin{equation}\label{eq:MomentumSpaceTimeCommutator1}
[\phi(p),q^+\bar{\phi}(q)]_{x^+=y^+}=-\delta^3(p+q)
\end{equation}
or
\begin{equation}\label{eq:MomentumSpaceTimeCommutator2}
[\phi(p),\bar{\phi}(q)]_{x^+=y^+}=-\frac{1}{q^+}\delta^3(p+q)
\end{equation}
Now, taking the Hamiltonian as a template, we want to generate transformations
\begin{equation}\label{eq:TemplateTransformation}
\delta_h^{(0)}\phi(p)=h\phi(p)\quad\quad\text{ with }\quad\quad h=\frac{p\bar{p}}{p^+}
\end{equation}
Then try
\begin{equation}\label{eq:FieldHamiltonian}
H=\int q^+dq^+d^2q\bar{\phi}(q)\frac{q\bar{q}}{q^+}\phi(q)
\end{equation}
and compute
\begin{equation}\label{eq:TryFieldTransformation}
\begin{split}
\delta_h^{(0)}\phi(p)&=[\phi(p),H]\\
&=\int q^+dq^+d^2q\Big([\phi(p),\bar{\phi}(q)]_{x^+=y^+}\Big)\frac{q\bar{q}}{q^+}\phi(q)\\
&=\int dq^+d^2q\Big([\phi(p),q^+\bar{\phi}(q)]_{x^+=y^+}\Big)\frac{q\bar{q}}{q^+}\phi(q)\\
&=\int dq^+d^2q\Big(-\delta^3(p+q)\Big)\frac{q\bar{q}}{q^+}\phi(q)\\
&=\frac{p\bar{p}}{p^+}\phi(-p)=\frac{p\bar{p}}{p^+}\phi(p)
\end{split}
\end{equation}
where the last equality follows from the complex fields being even functions of momenta. Hence, for any one Poincar{\'e} generator $g$, we define
\begin{equation}\label{eq:GeneralField Transformation}
\delta_g^{(0)}\phi(p)=[\phi(p),G]\quad\quad\text{ with }\quad\quad G=\int q^+dq^+d^2q\bar{\phi}(q)g\phi(q)
\end{equation}
The generalization to Fock fields is given by
\begin{equation}\label{eq:FreeHamiltonian}
G=\frac{1}{2}\int\gamma d\gamma dpd\bar{p}\langle\Phi\vert g\vert\Phi\rangle
\end{equation}
In an interacting field theory, the dynamical generators acquire non-linear contributions. The split $J=L+S$ then make little sense as the $L$ and $S$ part do not satisfy the algebra separately.

\subsection*{A note on the spin part of the dynamical Lorentz generators}
The underlying mechanical theory has three first class constraints
\begin{equation}\label{eq:FirstClassConstraints}
p^2\approx0\quad\quad\quad \alpha\cdot p\approx0\quad\quad\quad \alpha^\dagger\cdot p\approx0
\end{equation}
Solving the reparameterization constraint on the light-front $x^+\approx0$ yields the light-front mass-shell $p^-=p\bar p/\gamma$. Likewise, imposing $\alpha^+\approx0$ and $\alpha^{+\dagger}\approx0$, the gauge constraints can be solved
\begin{equation}\label{eq:SolvingGaugeConstraints}
\alpha^-=\frac{\alpha\bar p+\bar\alpha p}{\gamma}\quad\quad\quad\alpha^{-\dagger}=\frac{\alpha^\dagger\bar p+{\bar\alpha}^\dagger p}{\gamma}
\end{equation}
Then the spin parts of $j^-$ and $\bar j^-$ can then be calculated
\begin{align}\label{eq:DynamicalSpin}
s^-&=-i(\alpha^\dagger\alpha^--\alpha^{-\dagger}\alpha)=\frac{p}{\gamma}(\alpha^\dagger\bar\alpha-\bar\alpha^\dagger\alpha)=-i\frac{p}{\gamma}M\\
\bar s^-&=i(\bar\alpha^\dagger\alpha^--\alpha^{-\dagger}\bar\alpha)=\frac{\bar p}{\gamma}(-\alpha^\dagger\bar\alpha+\bar\alpha^\dagger\alpha)=i\frac{\bar p}{\gamma}M
\end{align}
This explains how the dynamical spin generators carry information about the gauge transformations.
\subsection{Structure of the algebra}\label{subsec:StructurePoincareAlgebra}
The four-dimensional Poincar{\'e} algebra consists of $45$ commutators, 22 of which are non-zero. It will be convenient to organize them in three groups, each containing three sub-types. This will facilitate the task to control them all. We will express the commutators in terms of the generators $g$ but the same equations hold for the $G$.

\paragraph{$\mathcal{K}-\mathcal{K}$ commutators}
  \begin{align*}
  [\mathcal{K},\mathcal{K}]&=\mathcal{K}\quad\quad \#9\\
  [\mathcal{K},\mathcal{K}]&=\mathcal{D}\quad\quad \#0\\
  [\mathcal{K},\mathcal{K}]&=\emptyset\quad\quad\; \#12\\
  \end{align*}

\noindent The non-zero commutators of this type are
\begin{equation}\label{eq:KKKCommutators}
\begin{split}
[\,j,p\;]&=-ip\hspace{60pt}[\,j,\bar{p}\;]=i\bar{p}\\
[\,j^+,\bar{p}\;]&=ip^+\hspace{53pt}[\,\bar{j}^+,p\;]=ip^+\\
[\,j^{+-},p^+\;]&=ip^+\\
[\,j,j^+\;]&=-ij^+\hspace{46pt}[\,j,\bar{j}^+\;]=i\bar{j}^+\\
[\,j^{+-},j^+\;]&=ij^+\hspace{41pt}[\,j^{+-},\bar{j}^+\;]=i\bar{j}^+
\end{split}
\end{equation}
Commuting two kinematic generators can never give a dynamic generator. This part of the algebra, being satisfied by the free theory by construction, therefore has no further consequences for the interactions. The three types of right hand sides of the commutators will be called {\it linear}, {\it non-linear} and {\it zero} respectively.

\paragraph{$\mathcal{K}-\mathcal{D}$ commutators}
  \begin{align*}
  [\mathcal{K},\mathcal{D}]&=\mathcal{K}\quad\quad \#6\\
  [\mathcal{K},\mathcal{D}]&=\mathcal{D}\quad\quad \#7\\
  [\mathcal{K},\mathcal{D}]&=\emptyset\quad\quad\; \#8\\
  \end{align*}

\noindent The non-zero commutators of the first subtype are
\begin{alignat}{3}\label{eq:KDKCommutators}
[\,j^-,p^+\;]&=-ip \quad\quad&[\,\bar{j}^-,p^+\;]&=-i\bar{p}\tag{$\cal{KD}$.1}\\
[\,j^+,h\;]&=-ip \quad\quad&[\,\bar{j}^+,h\;]&=-i\bar{p}\tag{$\cal{KD}$.2}\\
[\,j^+,\bar{j}^-\;]&=ij^{+-}-ij \quad\quad& [\,\bar{j}^+,j^-\;]&=ij^{+-}+ij\tag{$\cal{KD}$.3}
\end{alignat}
These commutators tell us that the kinematic transformations commute with the non-linear part of the dynamic transformations. In practice therefore, since the free part is satisfied by construction, they form a set of zero commutators together with the third subtype. These are
\begin{alignat}{3}\label{eq:KD0Commutators}
[\,p^+,h\;]&=0 \hspace{40pt}&[\,p,h\;]=0 \quad\quad\quad&[\,\bar{p},h\;]=0\tag{$\cal{KD}$.4}\\
[\,p,j^-\;]&=0 \hspace{20pt}&[\,\bar{p},\bar{j}^-\;]=0& \tag{$\cal{KD}$.5}\\
[\,j^+,j^-\;]&=0 \hspace{30pt}&[\,\bar{j}^+,\bar{j}^-\;]=0& \tag{$\cal{KD}$.6}\\
[\,j,h\;]&=0\tag{$\cal{KD}$.7}
\end{alignat}
Together these fix some of the structure of the interaction terms. The non-zero commutators of the second subtype are
\begin{alignat}{3}\label{eq:KDDCommutators}
[j^{+-},h]&=-ih\tag{$\cal{KD}$.8}\\
[\,j^-,\bar{p}\;]&=-ih\hspace{42pt}[\,\bar{j}^-,p\;]=-ih\tag{$\cal{KD}$.9}\\
[\,j^{+-},j^-\;]&=-ij^-\hspace{23pt}[\,j^{+-},\bar{j}^-\;]=-i\bar{j}^-\tag{$\cal{KD}$.10}\\
[\,j,j^-\;]&=-ij^-\hspace{38pt}[\,j,\bar{j}^-\;]=i\bar{j}^-\tag{$\cal{KD}$.11}
\end{alignat}
These work order by order in the interaction and fix still more of the structure. Taken together, the $\mathcal{K}-\mathcal{D}$ commutators determine the general form of the dynamical vertices up to $\gamma$-structure as will be explored in section \ref{subsec:RestrictionsKDAlgebra} through \ref{subsec:SummaryKinematicalConstraints}. 

\paragraph{$\mathcal{D}-\mathcal{D}$ commutators}
  \begin{align*}
  [\mathcal{D},\mathcal{D}]&=\mathcal{K}\quad\quad \#0\\
  [\mathcal{D},\mathcal{D}]&=\mathcal{D}\quad\quad \#0\\
  [\mathcal{D},\mathcal{D}]&=\emptyset\quad\quad\; \#3\\
  \end{align*}

\noindent With much of the structure already determined, the $\mathcal{D}-\mathcal{D}$ commutators yield recursive differential equations in the $\gamma_r$. Note that in super-Poincar{\'e} algebras there are $[\mathcal{D},\mathcal{D}]=\mathcal{D}$ commutators. In the non-supersymmetric case that is discussed here, there are only the zero commutators
\begin{equation}\label{eq:DDCommutators}
[\,h,j^-\;]=0\quad\quad\quad [\,h,\bar{j}^-\;]=0\quad\quad\quad [\,j^-,\bar{j}^-\;]=0
\end{equation}
The third commutator contain all information, as performing it requires the first two. There is a tentative analogy to be drawn here. Given that all interaction data will be carried by the deformations of $J^-$ and $\bar{J}^-$, the equation $[\,J^-,\bar{J}^-\;]=0$, although being a commutator, resembles the $\{Q,Q\}=0$ equation of deformed BRST-theory. Such deformations are known to result in strongly homotopy, or $L_\infty$ algebras, \cite{FulpLadaStasheff2002}. Some support for this connection comes from Siegel's and Zwiebach's work in the 1980's deriving the BRST gauge fixed string theory from light-front string theory by introducing ghost coordinates and constructing the BRST operator (and anti-BRST) out of the dynamical Lorentz operators \cite{SiegelZwiebach1987,AKHBNLinden1987}.

\section{Review of the cubic interaction}\label{sec:ReviewCubicInteraction}
Using the Hamiltonian as a template, free theory symmetry operators are defined as
\begin{equation}\label{eq:FreeHamiltonian}
H_{\scriptsize{(0)}}=\frac{1}{2}\int\gamma d\gamma dpd\bar{p}\langle\Phi\vert h\vert\Phi\rangle
\end{equation}
The cubic interaction can be written as
\begin{equation}\label{eq:CubicInteraction}
H_{\scriptsize{(1)}}=\frac{1}{3}\int\prod_{r=1}^3\gamma_rd\gamma_r dp_rd\bar{p}_r\langle\Phi_r\vert V_{123}\rangle
\end{equation}
where the cubic vertex operator $\vert V_{123}\rangle$ is defined by
\begin{equation}\label{eq:CubicVertex}
\begin{split}
\vert V_{123}\rangle&=\frac{g}{\kappa}\exp\Delta\vert0_{123}\rangle\Gamma^{-1}\delta({\textstyle\sum_r}\gamma_r)\delta({\textstyle\sum_r}p_r)\delta({\textstyle\sum_r}\bar{p}_r)\\
\Delta&=\kappa\sum_{r,s,t}Y^{rst}(\alpha_r^\dagger\alpha_s^\dagger\bar{\alpha}_t^\dagger\bar{\mathbb{P}}+\bar{\alpha}_r^\dagger\bar{\alpha}_s^\dagger\alpha_t^\dagger\mathbb{P})
\end{split}
\end{equation}
Here, $g$ is dimensionless coupling constant and $\kappa$ is of dimension $-1$. For the cubic interactions $g$ becomes the spin-1 coupling and $g\kappa$ becomes the spin-2 coupling. The higher spin coupling constants $g_\lambda$ come out as $g\kappa^{\lambda-1}$. The $Y^{rst}$ are rational functions of $\gamma$ to be determined by the algebra. $\Gamma$ is $\gamma_1\gamma_2\gamma_3$ and it compensates the measure factor. 

The dynamical Lorentz generators acquire the following cubic deformations
\begin{equation}\label{eq:CubicDynamicalLorentz}
\begin{split}
J^-_{(1)}&={\scriptstyle\frac{1}{3}}\int\prod_{r=1}^3\gamma_rd\gamma_r dp_rd\bar{p}_r\langle\Phi_r\vert\hat{x}_3\vert V_{123}\rangle\\
\bar{J}^-_{(1)}&={\scriptstyle\frac{1}{3}}\int\prod_{r=1}^3\gamma_rd\gamma_r dp_rd\bar{p}_r\langle\Phi_r\vert\hat{\bar{x}}_3\vert V_{123}\rangle\\
\end{split}
\end{equation}
These interaction terms deform the Poincar{\'e} algebra to first order, provided that
\begin{equation}\label{eq:CubicY}
Y^{rst}=\frac{\gamma_t}{\gamma_r\gamma_s}
\end{equation}
It should be clear how the momentum structure of \eqref{eq:CubicMomentumStructure} appears. The prefactors $\hat{x}_{(3)}$ are
\begin{equation}\label{eq:CubicXInsertions}
\hat{x}_{(3)}={\scriptstyle\frac{1}{3}}\sum_{r=1}^3x_r+{\scriptstyle\frac{2i}{3}}\kappa\sum_{r,s,t}Y^{rst}(\widetilde\gamma_r+\widetilde\gamma_s-\widetilde\gamma_t)\alpha_r^\dagger\alpha_s^\dagger\bar{\alpha}_t^\dagger
\end{equation}
with $\hat{\bar{x}}_{(3)}$ the c.c. of $\hat{x}_{(3)}$. This result, initially obtained in \cite{BBL1987}, will be rederived as a special case below.

\subsection*{A peculiarity}
In \cite{BBL1987} it was found that the vertex operator $\Delta$ needed terms with three oscillators and one momenta (as in \eqref{eq:CubicVertex}) in order to reproduce the known cubic interaction terms. Guided by string theory, it would have been thought that quadratic forms in oscillators and momenta would suffice, but that is not the case. At least not in four dimensions. As shown in \cite{BBL1987} and reproduced below, a string-like ansatz of the form
\begin{equation}\label{eq:CubicStringLike}
\Delta=\sum_{r,s}Y^{rs}\alpha_r^\dagger\bar{\alpha}_s^\dagger+\kappa\sum_{r}Y^{r}(\alpha_r^\dagger\bar{\mathbb{P}}+\bar{\alpha}_r^\dagger\mathbb{P})
\end{equation}
leads to the solution for the vertex functions $Y^{rs}$ and $Y^{r}$
\begin{equation}\label{eq:CubicStringLikeVertexFunctions}
Y^{rs}=\delta_{rs}\quad\quad\quad Y^{r}=\frac{1}{\gamma_r}
\end{equation}
Such a vertex is incapable of reproducing neither the Yang-Mills, nor the gravity, $3$-point couplings.

\subsection*{Deriving particular cubic interactions}
In this section we explain how to derive particular cubic interactions from the cubic vertex. Suppose we want a cubic interaction term involving three, possibly different, helicities $\lambda_1$, $\lambda_2$ and $\lambda_3$. The terms in the action are of the form $\phi_1\phi_2\phi_3$, $\bar\phi_1\phi_2\phi_3$, $\bar\phi_1\bar\phi_2\phi_3$ or $\bar\phi_1\bar\phi_2\bar\phi_3$ where the last two can be gotten from the first two using c.c. and cyclic symmetry in field labels.

For instance, to extract an interaction of the form $\bar\phi_1\phi_2\phi_3$, we chose a bra state excited by $\bar\phi_1\phi_2\phi_3\alpha^{\lambda_1}_1\bar\alpha^{\lambda_2}_2\bar\alpha^{\lambda_3}_3$ and insert it into \eqref{eq:CubicInteraction} using \eqref{eq:CubicVertex}. The matrix element to compute is
\begin{equation}
\bar\phi_1\phi_2\phi_3\langle0\vert\alpha_1^{\lambda_1}\bar\alpha_2^{\lambda_2}\bar\alpha_3^{\lambda_3}\exp\Delta\vert 0_{123}\rangle
\end{equation}
The annihilators will saturate any combination of creators $({\bar\alpha_1}^\dagger)^{\lambda_1}(\alpha_2^\dagger)^{\lambda_2}(\alpha_3^\dagger)^{\lambda_3}$ that appear in the expansion of $\exp\Delta$.

\subsubsection*{Yang-Mills and gravity}
For the Yang-Mills cubic vertex we get the matrix element
\begin{equation}\label{eq:YMVertexDerivation1}
\begin{split}
&\bar\phi_1\phi_2\phi_3\langle0\vert\alpha_1\bar\alpha_2\bar\alpha_3 \exp\Delta\vert 0_{123}\rangle\\
=&\bar\phi_1\phi_2\phi_3\langle0\vert\alpha_1\bar\alpha_2\bar\alpha_3\exp\big(\kappa Y^{rst}\alpha_r^\dagger\alpha_s^\dagger{\bar\alpha}_t^\dagger\bar{\mathbb P}+c.c.\big)\vert0_{123}\rangle\\
=&\kappa\bar\phi_1\phi_2\phi_3\big(Y^{231}+Y^{321}\big)\bar{\mathbb P}=2\kappa\bar\phi_1\phi_2\phi_3\frac{\gamma_1}{\gamma_2\gamma_3}\bar{\mathbb P}
\end{split}
\end{equation}
and the interaction Hamiltonian becomes
\begin{equation}\label{eq:YMVertexDerivation2}
\begin{split}
\frac{2\kappa}{3}\frac{g}{\kappa}\int\prod_{r=1}^3\gamma_rd\gamma_r dp_rd\bar{p}_r\bar\phi_1\phi_2\phi_3\frac{\gamma_1}{\gamma_2\gamma_3}\bar{\mathbb P}\Gamma^{-1}\delta({\textstyle\sum_r}\gamma_r)\delta({\textstyle\sum_r}p_r)\delta({\textstyle\sum_r}\bar{p}_r)\\
=-\frac{2g}{3}\int\prod_{r=1}^3d\gamma_r dp_rd\bar{p}_r\gamma_1\big(\frac{\bar p_2}{\gamma_2}-\frac{\bar p_3}{\gamma_3}\big)\bar\phi_1\phi_2\phi_3\delta({\textstyle\sum_r}\gamma_r)\delta({\textstyle\sum_r}p_r)\delta({\textstyle\sum_r}\bar{p}_r)
\end{split}
\end{equation}
Upon Fourier transforming back to configuration space we get
\begin{equation}\label{eq:YMVertexDerivation3}
-\frac{2g}{3}(2\pi)^{3/2}(-i)^2\int d^3x\partial^+\bar\phi\Big(\frac{\bar\partial}{\partial^+}\phi\phi-\phi\frac{\bar\partial}{\partial^+}\phi\Big)
\end{equation}
The need for anti-symmetrization is seen.

The numerical factor is partly conventional and not fixed by a purely cubic calculation. For instance, factors $(2\pi)^{2/3(\nu-2)}$ that result from Fourier transforming an $\nu$-vertex from momentum space to configuration space can be absorbed into the coupling $\sim g^{(\nu-2)}$.

Deriving the gravity cubic vertex is not much more work. The matrix element to compute is
\begin{equation}\label{eq:GravityVertexDerivation1}
\begin{split}
&\bar\phi_1\phi_2\phi_3\langle0\vert\alpha_1^2{\bar\alpha_2}^2{\bar\alpha_3}^2\exp\big(\kappa Y^{rst}\alpha_r^\dagger\alpha_s^\dagger{\bar\alpha}_t^\dagger\bar{\mathbb P}+c.c.\big)\vert0_{123}\rangle\\
=&\kappa^2\bar\phi_1\phi_2\phi_3\big(Y^{231}+Y^{321}\big)^2\bar{\mathbb P}^2=4\kappa^2\bar\phi_1\phi_2\phi_3\big(\frac{\gamma_1}{\gamma_2\gamma_3}\bar{\mathbb P}\big)^2
\end{split}
\end{equation}
The interaction Hamiltonian becomes
\begin{equation}\label{eq:YMVertexDerivation2}
\begin{split}
\frac{4g\kappa}{3}\int\prod_{r=1}^3d\gamma_r dp_rd\bar{p}_r\gamma_1^2\Big(\frac{\bar p_2}{\gamma_2}-\frac{\bar p_3}{\gamma_3}\Big)^2\bar\phi_1\phi_2\phi_3\delta({\textstyle\sum_r}\gamma_r)\delta({\textstyle\sum_r}p_r)\delta({\textstyle\sum_r}\bar{p}_r)
\end{split}
\end{equation}
and upon Fourier transforming back to configuration space we get
\begin{equation}\label{eq:YMVertexDerivation3}
\frac{8g\kappa}{3}(2\pi)^{3/2}\int d^3x(\partial^+)^2\bar\phi\Big[\Big(\frac{\bar\partial}{\partial^+}\Big)^2\phi\phi-\frac{\bar\partial}{\partial^+}\phi\frac{\bar\partial}{\partial^+}\phi\Big)\Big]
\end{equation}
%
\subsubsection*{Non-minimal coupling of spin $4$ to gravity}
Using both $Y^r$ and $Y^{rst}$ in $\Delta$, non-minimal coupling of higher spin to gravity can be derived. An example of a $\bar2-4-4$ vertex has field combination $\bar\phi_{(2)}\phi_{(4)}\phi_{(4)}$. Saturating the vertex with the bra Fock state $\bar\phi_1\phi_2\phi_3\alpha_1^2\bar\alpha_2^4\bar\alpha_3^4$ results in a cubic vertex with momentum structure
\begin{equation}\label{eq:GravityVertexDerivation1}
\begin{split}
&\bar\phi_1\phi_2\phi_3\langle0\vert\alpha_1^2\bar\alpha_2^4\bar\alpha_3^4\exp\big(\kappa Y^r\alpha_r^\dagger\bar{\mathbb P}+\kappa Y^{rst}\alpha_r^\dagger\alpha_s^\dagger{\bar\alpha}_t^\dagger\bar{\mathbb P}+c.c.\big)\vert0_{123}\rangle\\
\sim&\kappa^6\bar\phi_1\phi_2\phi_3\frac{\gamma_1^2}{\gamma_2^4\gamma_3^4}\bar{\mathbb{P}}^6\sim\kappa^6\gamma_1^2\gamma_2^2\gamma_3^2\Big(\frac{\bar p_2}{\gamma_2}-\frac{\bar p_3}{\gamma_3}\Big)^6\bar\phi_1\phi_2\phi_3
\end{split}
\end{equation}
where combinatorial factors are left out. In configuration space this transforms to a binomial form
\begin{equation}\label{eq:YMVertexDerivation3}
\sim\kappa^5\int d^3x\sum_{n=0}^6(-1)^n{6\choose n}(\partial^+)^2\bar\phi_{(2)}\Big[\frac{\bar\partial}{\partial^+}\Big]^{6-n}(\partial^+)^2\phi_{(4)}\Big[\frac{\bar\partial}{\partial^+}\Big]^{n}(\partial^+)^2\phi_{(4)}
\end{equation}
Incidentally, the transverse derivative structure is the same as in the two-loop pure spin-two off-shell light-front counterterm \cite{LBSSKunpubl1}. This hints at interesting quantum effects of higher helicity fields.

\paragraph{A comment} Already from the 1983 work on light-front cubic higher spin interactions it was clear that interactions between spin 2 and higher spin (at least of even spin) was not necessarily of the minimal form (i.e. with two derivatives) but rather governed by the higher spin fields \cite{BBB1983b,AKHB2004Thesis}. The interpretation and consequences of this remain to be understood.

\section{Framework for higher orders in the interaction}\label{sec:Framework}
\subsection{Vertices and generators}\label{subsec:VerticesGenerators}
Guided by the cubic interaction a framework for a formal expansion to all orders can be set up. The fields of equation \eqref{eq:FockSpaceField} and the free Hamiltonian \eqref{eq:FreeHamiltonian} remain the same. The interaction of order $\nu$ is written as
\begin{equation}\label{eq:OrderNuIntercationFull}
H_{\scriptsize{(\nu-2)}}=\frac{1}{\nu}\int\prod_{r=1}^\nu \gamma_rd\gamma_r dp_rd\bar{p}_r\langle\Phi_r\vert V_{1\ldots\nu}\rangle
\end{equation}
where the $\nu$-th order vertex is
\begin{equation}\label{eq:NuVertexFull}
\vert V_{1\ldots\nu}\rangle=\big(\frac{g}{\kappa}\big)^{4-\nu}\exp\Delta_\nu\vert0_{1\ldots\nu}\rangle\Gamma_\nu^{-1}\delta({\textstyle\sum_r}\gamma_r)\delta({\textstyle\sum_r}p_r)\delta({\textstyle\sum_r}\bar{p}_r)
\end{equation}
The power of the coupling is determined by dimensional analysis. Since the Hamiltonian must have dimension $1$, it follows that the vertex operator must have dimension determined by $1=4\nu-2\nu+\mathrm{dim}(\vert V_{1\ldots\nu}\rangle)$, that is, $\mathrm{dim}(\vert V_{1\ldots\nu}\rangle)=1-2\nu$. The three delta functions together with $\Gamma_\nu^{-1}$ carry dimension $-3-\nu$, hence the factor $\kappa^{\nu-4}$. Note that $\Gamma_\nu=\gamma_1\gamma_2\cdots\gamma_\nu$.

The choice of $\prod_{r=1}^\nu \gamma_rd\gamma_r$ integration measure and the concomitant need for the factor $\Gamma_\nu^{-1}$ is related to the field commutators of \eqref{eq:MomentumSpaceTimeCommutator2}, the measure essentially compensating for the $1/\gamma$ factors coming out of the commutators.

To clean up notation, the momentum integrations will be considered to be included as part of the Fock space inner product $\langle\,\vert\,\rangle$. Also, the momentum delta functions, the factor $\Gamma_\nu^{-1}$ as well as the coupling constants will be included in the vacua. This gives shorthand expressions, where we also give the dynamical Lorentz generators
\begin{equation}\label{eq:OrderNuIntercationShort}
\begin{split}
H_{\scriptsize{(\nu-2)}}&=\frac{1}{\nu}\prod_{r=1}^\nu\langle\Phi_r\vert V_{1\ldots\nu}\rangle\\
J^-_{\scriptsize{(\nu-2)}}&=\frac{1}{\nu}\prod_{r=1}^\nu\langle\Phi_r\vert\hat{x}_{(\nu)}\vert V_{1\ldots\nu}\rangle\\
\bar{J}^-_{\scriptsize{(\nu-2)}}&=\frac{1}{\nu}\prod_{r=1}^\nu\langle\Phi_r\vert\hat{\bar{x}}_{(\nu)}\vert V_{1\ldots\nu}\rangle\\
\vert V_{1\ldots\nu}\rangle&=\exp\Delta_\nu\vert\varnothing_{1\ldots\nu}\rangle\\
\vert\varnothing_{1\ldots\nu}\rangle&=\big(\frac{g}{\kappa}\big)^{4-\nu}\Gamma_\nu^{-1}\delta({\textstyle\sum_r}\gamma_r)\delta({\textstyle\sum_r}p_r)\delta({\textstyle\sum_r}\bar{p}_r)\vert0_{1\ldots\nu}\rangle
\end{split}
\end{equation}
Interaction data will be encoded by the $\Delta_\nu$ and the $\hat{x}_{(\nu)}$ and $\hat{\bar{x}}_{(\nu)}$. The rest of the formalism is essentially just scaffolding.

\subsection{Transformations and Fock field commutators}\label{subsec:TransformationsFockFieldCommutators}
The dynamical generators are now given by
\begin{equation}\label{eq:AllOrdersHamiltonian}
H=H_{(0)}+\sum_{\nu=3}^{\infty} H_{(\nu-2)}
\end{equation}
and similar expressions for the dynamical Lorentz generators $J^-$ and $\bar{J}^-$. The $\mathcal{D}-\mathcal{D}$ part of the Poincar{\'e} algebra then yields recursive equations for $\Delta_\nu$ and the $\hat{x}_\nu$ and $\hat{\bar{x}}_\nu$. In trying to solve these equations we clearly need an ansatz. The form of such an ansatz is restricted by the $\mathcal{K}-\mathcal{D}$ part of the algebra. Before tackling that, some more formalism must be set up. 

First of all, we have an option either to work directly in terms of the commutators such as $[H,J^-]$ or in terms of transformations such as $[\delta_H,\delta_{J^-}]\vert\Phi\rangle$. We will choose the second option although they are simply related through $[\delta_H,\delta_{J^-}]\vert\Phi\rangle=-\delta_{[H,J^-]}\vert\Phi\rangle$. In both options though, we have to sort out Fock space equal time commutators such as $[\vert\Phi_r\rangle,\vert\Phi_s\rangle]$. A generic transformation is given by
\begin{equation}\label{eq:GenericTransformation}
\delta_G\vert\Phi\rangle=[\vert\Phi\rangle,G]
\end{equation}
or
\begin{equation}\label{eq:GenericTransformation}
\delta_G^{(\mu)}\vert\Phi\rangle=[\vert\Phi\rangle,G_{(\mu)}]
\end{equation}
if a particular order of interaction $\mu$ is focused. What we need are commutators of the form
\begin{equation}\label{eq:TentativeFockFieldCommutators}
\begin{split}
[\vert\Phi_r\rangle,\vert\Phi_s\rangle]=\vert I_{rs}\rangle\\
[\langle\Phi_r\vert,\vert\Phi_s\rangle]=\langle I_{rs}\rangle\\
[\langle\Phi_r\vert,\langle\Phi_s\vert]=\langle I_{rs}\vert
\end{split}
\end{equation}
where we think of the $I_{rs}$ as {\it delta vertices} identifying momenta and Fock spaces labeled by $r$ and $s$. They are supposed to work in the following way
\begin{equation}\label{eq:WorkingsOfDeltaVertices}
\langle\Phi_r\vert\vert I_{rs}\rangle=\vert\Phi_s\rangle
\end{equation}
where we understand that momenta are identified through $p_r+p_s=0$ and the bra Fock space labeled by $r$ is turned into the ket Fock space labeled by $s$. The rest of the commutators of \eqref{eq:TentativeFockFieldCommutators} are interpreted correspondingly. Granted the existence of these delta vertices (they will be constructed in section \ref{subsec:ConstructionDeltaVertices}), we can derive the structure of the algebra. 

\subsection{The D - D algebra}\label{subsec:DDAlgebra}
For two generic dynamical generators $A$ and $B$ and a field $\vert\Phi_\chi\rangle$ we have
\begin{equation}\label{eq:AbstractAlgebra}
[\delta_A,\delta_B]\vert\Phi_\chi\rangle=0
\end{equation}
Expanding this equation a few orders in the interaction we get 
\begin{equation}\label{eq:ExpansionOfAlgebraToQuintic}
\begin{split}
&\text{Free:}\hspace{25pt}[\delta_A^{(0)},\delta_B^{(0)}]\vert\Phi_\chi\rangle=0\\
&\text{Cubic:}\hspace{17pt}\big([\delta_A^{(0)},\delta_B^{(1)}]+[\delta_A^{(1)},\delta_B^{(0)}]\big)\vert\Phi_\chi\rangle=0\\
&\text{Quartic:}\hspace{9pt}\big([\delta_A^{(0)},\delta_B^{(2)}]+[\delta_A^{(2)},\delta_B^{(0)}]\big)\vert\Phi_\chi\rangle=-[\delta_A^{(1)},\delta_B^{(1)}]\vert\Phi_\chi\rangle\\
&\text{Quintic:}\hspace{9pt}\big([\delta_A^{(0)},\delta_B^{(3)}]+[\delta_A^{(3)},\delta_B^{(0)}]\big)\vert\Phi_\chi\rangle=-[\delta_A^{(1)},\delta_B^{(2)}]+[\delta_A^{(2)},\delta_B^{(1)}]\vert\Phi_\chi\rangle\\
\end{split}
\end{equation}
The recursive nature of the equations is apparent, where for instance in the third line, the cubic transformations act as ''sources'' to the quartic and so on. Given that the transformations are generated by \eqref{eq:OrderNuIntercationShort} which can be thought of as a kind of higher order products of the fields, we see that the algebra is reminiscent of a $L_\infty$ algebra \cite{LadaStasheff1993a}. Note that in a supersymmetric theory the algebra gets modified due to the presence of $[\mathcal{D},\mathcal{D}]=\mathcal{D}$ commutators.

The general form of these equations can be written as
\begin{equation}\label{eq:ExpansionOfAlgebraGeneral}
\big([\delta_A^{(0)},\delta_B^{(\nu)}]+[\delta_A^{(\nu)},\delta_B^{(0)}]\big)\vert\Phi_\chi\rangle=-\sum_{\mu=1}^{\nu-1}\Big([\delta_A^{(\mu)},\delta_B^{(\nu-\mu)}]+[\delta_A^{(\nu-\mu)},\delta_B^{(\mu)}]\Big)\vert\Phi_\chi\rangle
\end{equation}
The left hand sides of these equations will be called the {\it differential} commutator and the right hand side the {\it source} commutator. The differential can be further reduced to a concrete form suitable for calculation (see section \ref{subsec:ComputationCommutators}). We list the three differentials corresponding to the $\mathcal{D}-\mathcal{D}$ commutators \eqref{eq:DDCommutators}.
\begin{subequations}\label{eq:DifferentialSide}
\begin{equation}\label{eq:DifferentialSideA}
\sum_{r=1}^\nu j_r^-\vert V_{1\ldots\nu}\rangle-\sum_{r=1}^\nu h_r\hat{x}_{(\nu)r}\vert V_{1\ldots\nu}\rangle
\end{equation}
\begin{equation}\label{eq:DifferentialSideB}
\sum_{r=1}^\nu \bar{j}_r^-\vert V_{1\ldots\nu}\rangle-\sum_{r=1}^\nu h_r\hat{\bar{x}}_{(\nu)r}\vert V_{1\ldots\nu}\rangle
\end{equation}
\begin{equation}\label{eq:DifferentialSideC}
\sum_{r=1}^\nu \bar{j}_r^-\hat{x}_{(\nu)}\vert V_{1\ldots\nu}\rangle-\sum_{r=1}^\nu j_r^-\hat{\bar{x}}_{(\nu)}\vert V_{1\ldots\nu}\rangle
\end{equation}
\end{subequations}
%

\subsection{Some restrictions from the K - D algebra}\label{subsec:RestrictionsKDAlgebra}
Let $k$ be a kinematic generator and $d$ a dynamic and $\hat{d}\,\vert V_\nu\rangle$ be short for its vertex form (so that $\hat{h}=1$, $\hat{j}^-=\hat{x}$ and $\hat{\bar{j}}^-=\hat{\bar{x}}$.) Then the $\mathcal{K}-\mathcal{D}$ commutators with zero or linear right hand side, can be written in a form suitable for direct calculation as
\begin{equation}\label{eq:KDVariations}
[\delta_k^{(0)},\delta_d^{(\nu)}]\vert\Phi\rangle=0\quad\Rightarrow\quad\sum_{r=1}^\nu k_r\hat{d}\,\vert V_{1\ldots\nu}\rangle=0
\end{equation}
A few of the $\mathcal{K}-\mathcal{D}$ commutators can now be discarded off at once.
\begin{itemize}
  \item  The commutators of ($\cal{KD}$.4) tell us that there are no explicit occurrences of $x^-$, $x$ or $\bar{x}$ in $\vert V_{1\ldots\nu}\rangle$. The zero sums of the $\gamma,p,\bar{p}$ momenta are guaranteed by the delta functions.
  \item The commutators of ($\cal{KD}$.1) tell us that the prefactors $\hat{x}$ and $\hat{\bar{x}}$ do not contain $x^-$.
  \item The commutators of ($\cal{KD}$.5) and ($\cal{KD}$.9) tell us that the prefactor $\hat{x}$ contain the coordinate $x$ linearly (but not $\bar{x}$) and correspondingly that the prefactor $\hat{\bar{x}}$ contain the coordinate $\bar{x}$ linearly (but not $x$).
\end{itemize}
These results should be intuitively clear from the construction of the light-front frame. Sharper restrictions on the vertices will follow from the ($\cal{KD}$.2), ($\cal{KD}$.7) and ($\cal{KD}$.8) commutators. 

\subsection{Ansatz for the vertex functions}\label{subsec:AnsatzDelta}
Much of the simplicity of the cubic vertex is due to the nice factorization of $\Delta_3$ into the factor $Y^{rst}\alpha_r^\dagger\alpha_s^\dagger\bar{\alpha}_t^\dagger$ and the special transverse momentum factor $\bar{\mathbb{P}}$. As we will see, this form is fixed by the $\mathcal{K}-\mathcal{D}$ commutators. Clearly it is essential to check if this structure generalize to higher order vertices.

The investigation will be organized by doing the general $\vert V_{1\ldots\nu}\rangle$ and the known $\vert V_{123}\rangle$ in parallel. Suppose we weren't quite sure about the factorization property of $\Delta_3$. A slightly more general expression would be
\begin{equation}\label{eq:MoreGeneralDelta3}
\Delta_3=Y^{rstu}\alpha_r^\dagger\alpha_s^\dagger\bar{\alpha}_t^\dagger\bar{p}_u+c.c.
\end{equation}
where sums over the labels are understood (c.c. stands for the complex conjugated term). Apart from the balance between bared and unbared oscillators and momenta, this is almost as general as it can be, all $\gamma$ dependence maintained by the $Y$. It might seem that we are excluding possibilities such as
\begin{equation}\label{eq:NotNeededDeltaNu}
Y^{rstuv}\alpha_r^\dagger\alpha_s^\dagger\bar{\alpha}_t^\dagger\bar{p}_u\gamma_v+c.c.
\end{equation}
However, such terms can taken care of by a redefinition of the $Y$ according to $Y^{rstu}=Y^{rstuv}\gamma_v$.

Based on these observations, we try contributions to $\Delta_\nu$ of the form
\begin{equation}\label{eq:AnsatzDeltaNuFull}
Y^{r_1\ldots r_k s_1\ldots s_l t_1\ldots t_m u_1\ldots u_n}\alpha_{r_1}^\dagger\ldots\alpha_{r_k}^\dagger\bar{\alpha}_{s_1}^\dagger\ldots\bar{\alpha}_{s_l}^\dagger p_{t_1}\ldots p_{t_m}\bar{p}_{u_1}\ldots \bar{p}_{u_n}
\end{equation}
where summations are understood and the c.c. should be added. The $Y$ are symmetric in the $r,s,t,u$ labels separately. We abbreviate it to\footnote{Dimensional factors of $\kappa$ are suppressed.}
\begin{equation}\label{eq:AnsatzDeltaNuAbbreviated}
Y^{(k)(l)(m)(n)}(\alpha^\dagger)^k(\bar{\alpha}^\dagger)^lp^m\bar{p}^n
\end{equation}
only showing explicit labels when needed. For any such term in $\Delta_\nu$ there is a complex conjugated mirror term. We defer discussing this to section \ref{sec:ComputationDDDifferential}. Until then, $\Delta_\nu$ will denote any term of the form \eqref{eq:AnsatzDeltaNuAbbreviated}.

\subsubsection*{Further restrictions from the K - D algebra}
We start with the commutator of ($\cal{KD}$.7).
\begin{equation}\label{eq:KD7Calculation}
\begin{split}
\sum_{r=1}^\nu j_r\vert V_\nu\rangle&=\sum_{r=1}^\nu\big(x_r\bar{p}_r-\bar{x}_rp_r-i(\alpha_r^\dagger\bar{\alpha}_r-\bar{\alpha}_r^\dagger\alpha_r)\big)\exp\Delta_\nu\vert\varnothing_{1\ldots\nu}
\rangle\\
&=\big(i(n-m)-i(k-l)\big)\Delta_\nu\exp\Delta_\nu\vert\varnothing_{1\ldots\nu}\rangle=0
\end{split}
\end{equation}
To satisfy this we require $(n-m)=(k-l)$ which can be interpreted as a balance between transverse orbital angular momentum and helicity. Written as $k+m=l+n$ the requirement also means that the overall number of unbared and bared $\alpha$'s and $p$'s should be the same. It is satisfied by $\Delta_3$. 

We then turn to the commutators of ($\cal{KD}$.2). Here we will do it first for the cubic case
\begin{equation}\label{eq:KD2CalculationCubic}
\begin{split}
\sum_{r=1}^3 j^+\vert V_3\rangle&=\sum_{r=1}^3 x_r\gamma_r\exp\Delta_3\vert\varnothing_{123}
\rangle\\
&=\exp\Delta_3\sum_{r=1}^3 x_r\gamma_rY^{rstu}\alpha_r^\dagger\alpha_s^\dagger\bar{\alpha}_t^\dagger\bar{p}_u\vert\varnothing_{123}\rangle\\
&=i\exp\Delta_3Y^{rstu}\alpha_r^\dagger\alpha_s^\dagger\bar{\alpha}_t^\dagger\gamma_u\vert\varnothing_{123}\rangle=0
\end{split}
\end{equation}
To satisfy this equation we need
\begin{equation}\label{eq:RequirementCubicJPlus}
\sum_{u=1}^3Y^{rstu}\gamma_u=0\quad\text{ for all combinations of $r,s,t$}
\end{equation}
For the cubic vertex, there is a nice re-summation formula.\footnote{I don't know the original reference. I learned it from N. Linden.} It states that for any indeterminate quantities $c_r$
\begin{equation}\label{eq:ResummationFormulaCubic}
\sum_{r=1}^3 c_rp_r=\frac{1}{3}\Big(\sum_{r=1}^3 c_r\gamma_r\Big)\Big(\sum_{s=1}^3\frac{p_s}{\gamma_s}\Big)+\frac{1}{3\Gamma}\Big(\sum_{r=1}^3 c_r\gamma_r\widetilde{\gamma}_r\Big)\mathbb{P}
\end{equation}
where $\Gamma=\gamma_1\gamma_2\gamma_3$. It can be proved by direct calculation using momentum conservation and cyclic symmetry in the labels.

If the re-summation formula is applied to $Y^{rstu}\bar{p}_u$, that is with $c_u=Y^{rstu}$ we get
\begin{equation}\label{eq:ResummationYCubic}
\sum_{u=1}^3 Y^{rstu}\bar{p}_u=\frac{1}{3}\Big(\sum_{u=1}^3 Y^{rstu}\gamma_u\Big)\Big(\sum_{s=1}^3\frac{\bar{p}_s}{\gamma_s}\Big)+\frac{1}{3\Gamma}\Big(\sum_{u=1}^3 Y^{rstu}\gamma_u\widetilde{\gamma}_u\Big)\bar{\mathbb{P}}
\end{equation}
The first term on the right is zero by \eqref{eq:RequirementCubicJPlus} and the factor in front of $\bar{\mathbb{P}}$ amounts to a redefinition $Y^{rst}=1/(3\Gamma)\sum_{u=1}^3 Y^{rstu}\gamma_u\widetilde{\gamma}_u$. This is the way the form of the cubic $\Delta_3$ of equation \eqref{eq:CubicVertex} emerges. The resummation formula \eqref{eq:ResummationFormulaCubic} can be generalized to arbitrary vertex order \cite{AKHB2016a}. For vertex order $\nu$ there are $\nu-2$ independent $\mathbb{P}_{ij}$ and an ansatz has to be chosen in terms of such $\mathbb{P}_{ij}$ for each particular vertex order. For the quartic vertex, see \cite{AKHB2016c}. Here we will work with the form of equation \eqref{eq:AnsatzDeltaNuFull} which is general for all vertices.

In analogy to \eqref{eq:RequirementCubicJPlus} we get when applying $j^+$ and $\bar{j}^+$ to $\vert V_{1\ldots\nu}\rangle$
\begin{equation}\label{eq:GammaSlotCondition}
\begin{split}
Y^{(k)(l)t_1\ldots t_i\ldots t_m (n)}\gamma_{t_i}=0\\
Y^{(k)(l)(n)u_1\ldots u_i\ldots u_n }\gamma_{u_i}=0
\end{split}
\end{equation}
meaning that replacing any one transverse momentum $p$ or $\bar{p}$ with $\gamma$ and summing over labels yield zero (irrespective of all the other labels).

Next we impose the commutator of ($\cal{KD}$.8). Commutators involving the operator $j^{+-}$ require some care. The computational form of the commutator $[\delta_{j^{+-}},\delta_h]\phi=i\delta_h\phi$ is (see section \ref{subsec:ComputationCommutators})
\begin{equation}\label{eq:JPlusMinusCom1}
-\sum_{r=1}^\nu\big(i\gamma_r\frac{\partial}{\partial\gamma_r}+i\big)|V_\nu\rangle=i|V_\nu\rangle
\end{equation}
The $\gamma$ dependence of the vertex sits in three places: $\Delta_\nu$, $\Gamma_\nu^{-1}$ and the $\gamma$ conservation delta function. For the latter two we have
\begin{align}\label{eq:HomogenietyVacuum}
\sum_{r=1}^\nu \gamma_r\frac{\partial}{\partial\gamma_r}\Gamma_\nu^{-1}&=-\nu\Gamma_\nu^{-1}\\
\sum_{r=1}^\nu \gamma_r\frac{\partial}{\partial\gamma_r}\delta\big(\sum_{s=1}^\nu \gamma_s\big)&=-\delta\big(\sum_{s=1}^\nu \gamma_s\big)
\end{align}
Therefore, equation \eqref{eq:JPlusMinusCom1}, becomes
\begin{equation}\label{eq:JPlusMinusCom2}
-\Big(\sum_{r=1}^\nu i\gamma_r\frac{\partial\Delta_\nu}{\partial\gamma_r}+i\nu-i\nu-i\Big)\exp\Delta_\nu\vert\varnothing_{1\ldots\nu}\rangle=i|V_\nu\rangle
\end{equation}
that is
\begin{equation}\label{eq:JPlusMinusCom3}
\sum_{r=1}^\nu \gamma_r\frac{\partial\Delta_\nu}{\partial\gamma_r}=0
\end{equation}
telling us that the functions $Y$ making up $\Delta$ are homogeneous functions of the $\gamma_r$ of degree zero. Incidentally this also shows that the $Y$-functions in $\Delta$ comes with a dimensional factor of $\kappa^{m+n}$. These factors will be suppressed.
\subsection{Ans{\"a}tze for the dynamical Lorentz prefactors}\label{subsec:AnsatzPrefactors}
A few more restrictions on the prefactors follow from the remaining $\mathcal{K}-\mathcal{D}$ commutators involving two Lorentz operators. These are as follows.

\begin{itemize}
  \item  The commutators of ($\cal{KD}$.3) and ($\cal{KD}$.6) tell us that when any one transverse momenta in $\hat{x}_{(\nu)}$ and $\hat{\bar{x}}_{(\nu)}$ are replaced by a $\gamma$ the result is zero. This is the same kind of condition as \eqref{eq:GammaSlotCondition} on the $Y$. For instance, $\sum_r x_r\gamma_r \hat{\bar{x}}\vert V\rangle=\sum_r[x_r\gamma_r,\hat{\bar{x}}]\vert V\rangle=0$ using ($\cal{KD}$.2).
  \item The commutator of ($\cal{KD}$.10) fixes the $\gamma$ homogeneity of the prefactors to zero. A short calculation yields
 
 $$\sum_r(i\gamma_r\frac{\partial}{\partial\gamma_r}+i)\hat{x}\vert V\rangle=\sum_r(i\gamma_r\frac{\partial}{\partial\gamma_r}\hat{x})\vert V\rangle+\hat{x}\sum_r(i\gamma_r\frac{\partial}{\partial\gamma_r}+i)\vert V\rangle$$

The second term equals $-i\hat{x}\vert V\rangle$ by \eqref{eq:JPlusMinusCom1}, hence the first term must be zero, and we get the homogeneity requirement
\begin{equation}\label{eq:HomogeneityPrefactors}
\sum_r \gamma_r\frac{\partial}{\partial\gamma_r}\hat{x}=0
\end{equation}
A similar equation holds for $\hat{\bar{x}}$.

  \item Finally the commutators of ($\cal{KD}$.11) fix the balance between unbared and bared momenta and oscillators. A calculation similar to \eqref{eq:KD7Calculation} yields $(k^\prime-l^\prime)-(n^\prime-m^\prime)=1$ where $k^\prime$, $l^\prime$, $m^\prime$ and $n^\prime$ are the numbers of $\alpha$, $\bar\alpha$, $p$ and $\bar{p}$ respectively. The ans{\"a}tze for the prefactors are
\begin{equation}\label{eq:AnsatzPrefactors}
\begin{split}
\hat{x}=a^r x_r+c^{(k^\prime)(l^\prime)(m^\prime)(n^\prime)}(\alpha^\dagger)^{k^\prime}(\bar{\alpha}^\dagger)^{l^\prime}p^{m^\prime}\bar{p}^{n^\prime}\\
\hat{\bar{x}}=\bar{a}^r \bar{x}_r+\bar{c}^{(k^\prime)(l^\prime)(m^\prime)(n^\prime)}(\bar{\alpha}^\dagger)^k{^\prime}(\alpha^\dagger)^{l^\prime}\bar{p}^{m^\prime}p^{n^\prime}
\end{split}
\end{equation}
The balance condition $(k^\prime-l^\prime)-(n^\prime-m^\prime)=1$ is entirely natural, it just says that the second term must have the same  total transverse orbital and spin angular momentum as the coordinate. This balance condition can be satisfied by $k^\prime=k$, $l^\prime=l$, $m^\prime=m$ and $n^\prime=n-1$ and this is indeed what follows from the dynamical part of the algebra.

The linear terms $a^r x_r$ and $\bar{a}^r \bar{x}_r$ correspond to the unity term in the expansion of $\exp\Delta$. Each $c$ or $\bar c$ term correspond precisely to the $Y$-function with the same oscillator basis. Thus $a^r x_r$ and $\bar{a}^r \bar{x}_r$ is what would remain of the $c$ or $\bar c$ terms if all the $Y$ functions were zero (corresponding to a theory with only spin $0$ fields). The solutions are $a_r={\bar a}_r=1/\nu$.
\end{itemize}
With this we have now extracted all information from the $\mathcal{K}-\mathcal{D}$ commutators. It will summarized in the next section.

\subsection{Summary of kinematical constraints}\label{subsec:SummaryKinematicalConstraints}
Most of the kinematic constraints are quite trivial and just exclude or fix certain dependencies as we've seen in sections \ref{subsec:RestrictionsKDAlgebra} through \ref{subsec:AnsatzPrefactors}. There are however three requirements that come out of the analysis that are crucial. They are as follows.

\begin{description}
  \item[$\gamma$-homogeneity] This comes from the commutators with $j^{+-}$. 

	$Y$, $c$ and $\bar{c}$ are all homogeneous functions of $\gamma$ of degree $0$. 
  \item[$\gamma$-replacement property] This comes from the commutators with $j^{+}$.

	Replacing a transverse momentum in $Y$, $c$ and $\bar{c}$ with $\gamma$ gives zero. For the cubic this is implemented by the construction of the $\mathbb{P}$ and $\mathbb{\bar{P}}$. In configuration space this become binomial expansions.
  \item[Angular momentum balance] This comes from the commutators with $j$.

	The number of unbared and bared creators and transverse momenta in $Y$ must be the same. 

	The number of unbared and bared creators and transverse momenta in $c$ and $\bar{c}$ must differ by $\pm1$. 
\end{description}   

\section{Computation of the D - D differential}\label{sec:ComputationDDDifferential}
In this section we compute the differential \eqref{eq:DifferentialSideA} explicitly using the ans{\"a}tze \eqref{eq:AnsatzDeltaNuFull} and \eqref{eq:AnsatzPrefactors}. Since this is a crucial part it will be presented in some detail. To reduce clutter the order $\nu$ vertex is written $\vert V\rangle$ and explicit sums are supposed to run from $1$ to $\nu$. Repeated labels are summed over. There is a point of book-keeping to clarify the formulas to follow.

\subsection{Book-keeping}\label{subsec:BookKeeping}
The ansatz for $\Delta$ contains terms of the form
\begin{equation}\label{eq:TermsAndMirrorTerms1}
\Delta^{(kl)(mn)}=Y^{(k)(l)(m)(n)}(\alpha^\dagger)^k(\bar{\alpha}^\dagger)^lp^m\bar{p}^n
\end{equation}
The products of $k$ oscillators of type $\alpha_{r_i}^\dagger$ and $l $ oscillators of type $\bar{\alpha}_{s_i}^\dagger$ with indices ranging over the field labels, work as a basis. Terms with different numbers of oscillators never mix in the computation of the differential. However, for a certain number of oscillators $k$ and $l$, there must be a mirror term  $\Delta^{(lk)(nm)}$ with un-bared and bared oscillators and momenta interchanged. This mirror term is 
\begin{equation}\label{eq:TermsAndMirrorTerms2}
\Delta^{(lk)(nm)}=Y^{(l)(k)(n)(m)}(\alpha^\dagger)^l(\bar{\alpha}^\dagger)^kp^n\bar{p}^m
\end{equation}
and must be the complex conjugate of $\Delta^{(kl)(mn)}$. Reality then requires the $Y$-functions to be real and symmetric in the sense
\begin{equation}\label{eq:SymmetryY}
Y^{(l)(k)(n)(m)}=Y^{(k)(l)(m)(n)}
\end{equation}
When considering a particular ansatz $\Delta_\nu$ with a total number of oscillators $k+l$ it is convenient to take $k\geq l$. This is made explicit by introducing $\Delta_\nu^\prime$ and $\bar{\Delta}_\nu^\prime$ defined by
\begin{align}\label{eq:AnsatzDeltaPrimeNu}
\Delta_\nu^\prime&=Y^{(k)(l)(m)(n)}(\alpha^\dagger)^k(\bar{\alpha}^\dagger)^lp^m\bar{p}^n=\Delta^{(kl)(mn)}\quad\quad k\geq l\\
\bar{\Delta}_\nu^\prime&=Y^{(k)(l)(m)(n)}(\bar{\alpha}^\dagger)^k(\alpha^\dagger)^l\bar{p}^mp^n=\Delta^{(lk)(nm)}\quad\quad k\geq l
\end{align}
where \eqref{eq:SymmetryY} is taken into account. For any particular value for $k$ we then have
\begin{equation}\label{eq:DeltaEq1}
\Delta_\nu=\Delta_\nu^\prime+\bar{\Delta}_\nu^\prime\quad\quad\text{ when }k>l
\end{equation}
whereas when $k=l$ it is more convenient to take
\begin{equation}\label{eq:DeltaEq2}
\Delta_\nu=\frac{1}{2}(\Delta_\nu^\prime+\bar{\Delta}_\nu^\prime)
\end{equation}
since then $\Delta_\nu^\prime=\bar{\Delta}_\nu^\prime$.

The importance of this book-keeping has to do with the cubic dynamical Lorentz generators. The prefactors $\hat x$ and $\hat{\bar x}$ of equations \eqref{eq:JPlusMinusCom1} have similar formal expansions over the oscillator basis as $\Delta$ with two important differences: First, they are not real but conjugates of each other, and second, $\hat x$ contains one less factor of $\bar{\mathbb P}$ and $\hat{\bar x}$ one less factor of $\mathbb P$ as compared to the corresponding term in $\Delta$.

Now suppose we choose a set of numbers $k,l,m,n$ subject to $k-l=n-m$ and $k\geq l$ and proceed to compute the $[J^-,H]$ differential. We get two equations, one for $\Delta^\prime$ and another for the mirror term ${\bar\Delta}^\prime$. Both involve the same $Y$ function (due to the above noted symmetry as implied by the reality $H$) but the $c$ contributions are different as will become clear below.

\subsection{Computation of the first part of \eqref{eq:DifferentialSideA}}\label{subsec:ComputationJ-V}
\paragraph{Terms from $xh$ :}
These are
\begin{equation}\label{eq:CalculationJMinusHxh}
\begin{split}
\sum_r x_rh_r\vert V\rangle&=\sum_r\big(h_rx_r+[x_r,h_r]\big)\vert V\rangle\\
&=\sum_r h_r\big([x_r,e^\Delta]+e^\Delta{x_r}\big)\vert\varnothing\rangle+i\sum_r\frac{p_r}{\gamma_r}\vert V\rangle\\
&=\sum_r h_r[x_r,e^\Delta]\vert\varnothing\rangle+\sum_r e^\Delta h_rx_r\vert\varnothing\rangle+i\sum_r\frac{p_r}{\gamma_r}\vert V\rangle
\end{split}
\end{equation}
Here, the second and the third terms will cancel contributions from the other parts of the commutator. To understand the first term, simply note that $h_r[x_r,\cdot]$ will hit the each and every $\bar{p}$ in $\Delta$ replacing it by $i\bar{p}_rp_r/\gamma_r$. The first term therefore becomes
\begin{equation}\label{eq:JMinusHxh}
\begin{split}
&iY^{(k)(l)(m)u_1\ldots u_n}(\alpha^\dagger)^k(\bar{\alpha}^\dagger)^lp^m\bar{p}_{u_1}\ldots\bar{p}_{u_n}\bigg(\frac{p_{u_1}}{\gamma_{u_1}}+\ldots+\frac{p_{u_n}}{\gamma_{u_n}}\bigg)\vert V\rangle\\
+&iY^{(k)(l)t_1\ldots t_m(n)}(\bar{\alpha}^\dagger)^k(\alpha^\dagger)^l\bar{p}_{t_1}\ldots\bar{p}_{t_m}\bigg(\frac{p_{t_1}}{\gamma_{t_1}}+\ldots+\frac{p_{t_m}}{\gamma_{t_m}}\bigg)p^n\vert V\rangle
\end{split}
\end{equation}
Here we recognize the contributions from $\Delta_\nu^\prime$ and $\bar{\Delta}_\nu^\prime$ on the first and second lines respectively.

\paragraph{Terms from $ip\frac{\partial}{\partial\gamma}$ :}
These terms follow straightforwardly, the catch point being to note that
\begin{equation}\label{eq:DGammaOnVacuum}
\sum_r p_r\frac{\partial}{\partial\gamma_r}\Gamma^{-1}\delta({\textstyle\sum_r}\gamma_r)=-\sum_r\frac{p_r}{\gamma_r}\delta({\textstyle\sum_r}\gamma_r)
\end{equation}
Using this we get
\begin{equation}\label{eq:CalculationJMinusHpdgamma1}
\begin{split}
i\sum_r p_r\frac{\partial}{\partial\gamma_r}e^\Delta\vert \varnothing\rangle&=i\sum_r \bigg(p_r\frac{\partial\Delta}{\partial\gamma_r}\bigg)e^\Delta\vert \varnothing\rangle+i\sum_r e^\Delta p_r\frac{\partial}{\partial\gamma_r}\vert \varnothing\rangle\\
&=i\sum_r p_r\frac{\partial Y^{(k)(l)(m)(n)}}{{\partial\gamma_r}}(\alpha^\dagger)^k(\bar{\alpha}^\dagger)^lp^m\bar{p}^n\vert V\rangle\\
&+i\sum_r p_r\frac{\partial Y^{(k)(l)(m)(n)}}{{\partial\gamma_r}}(\bar{\alpha}^\dagger)^k(\alpha^\dagger)^l\bar{p}^m p^n\vert V\rangle\\
&-i\sum_r\frac{p_r}{\gamma_r}\vert V\rangle
\end{split}
\end{equation}
The third term cancels the third term from \eqref{eq:CalculationJMinusHxh}.

\paragraph{Terms from $-\frac{i}{\gamma}Mp$ :} The annihilators in $M$ act on the creators in $\Delta$ inserting a term $p/\gamma$ for every $\alpha^\dagger$ and a term $-p/\gamma$ for every $\bar{\alpha}^\dagger$. The result is
\begin{equation}\label{eq:JMinusHMp}
\begin{split}
&-iY^{r_1\ldots r_k s_1\ldots s_l(m)(n)}(\alpha^\dagger)^k(\bar{\alpha}^\dagger)^l\bigg(\frac{p_{r_1}}{\gamma_{r_1}}+\ldots+\frac{p_{r_k}}{\gamma_{r_k}}-\frac{p_{s_1}}{\gamma_{s_1}}-\ldots-\frac{p_{s_l}}{\gamma_{s_l}}\bigg)p^m\bar{p}^n\vert V\rangle\\
&+iY^{r_1\ldots r_k s_1\ldots s_l(m)(n)}(\bar{\alpha}^\dagger)^k(\alpha^\dagger)^l\bigg(\frac{p_{r_1}}{\gamma_{r_1}}+\ldots+\frac{p_{r_k}}{\gamma_{r_k}}-\frac{p_{s_1}}{\gamma_{s_1}}-\ldots-\frac{p_{s_l}}{\gamma_{s_l}}\bigg)\bar{p}^m p^n\vert V\rangle
\end{split}
\end{equation}

\subsection{Computation of the second part of \eqref{eq:DifferentialSideA}}\label{subsec:ComputationhxV}
This essentially entails commuting the prefactor $\hat{x}$ through $e^\Delta$. We do it first for the coordinate piece, then for the oscillator piece.

\paragraph{Terms from $a^r x_r$ :} The computation runs as follows
\begin{equation}\label{eq:CalculationJMinusHhax}
\begin{split}
-\sum_t h_t\bigg(\sum_r a_rx_r\bigg)\vert V\rangle)=&-\sum_t h_t e^\Delta\sum_r a_r[x_r,\Delta]\vert\varnothing\rangle\\
&-\sum_t e^\Delta h_t\sum_r a_rx_r\vert\varnothing\rangle
\end{split}
\end{equation}
The second term cancels the second term of \eqref{eq:CalculationJMinusHxh} provided we choose all $a_r=1/\nu$. This is because all $x_r$ are equal on the vacuum, a consequence of momentum conservation, or locality in transverse directions.

We are left with the first term. The commutator $[x_r,\Delta]$ appeared already in \eqref{eq:CalculationJMinusHxh}. Here it is ''scalar multiplied'' over the field labels into the numbers $a_r$ instead of the free Hamiltonian.
\begin{equation}\label{eq:XrDeltaCommutator}
\begin{split}
\sum_r a_r[x_r,\Delta]&=in\sum_r a_rY^{(k)(l)(m)ru_2\ldots u_n}(\alpha^\dagger)^k(\bar{\alpha}^\dagger)^lp^m\bar{p}_{u_2}\ldots\bar{p}_{u_n}\\
&+im\sum_r a_rY^{(k)(l)rt_2\ldots t_m(n)}(\bar{\alpha}^\dagger)^k(\alpha^\dagger)^l\bar{p}_{t_2}\ldots\bar{p}_{t_m}p^n
\end{split}
\end{equation}
and the surviving terms are
\begin{equation}\label{eq:JMinusHhax}
\begin{split}
&-in\sum_r a_rY^{(k)(l)(m)ru_2\ldots u_n}(\alpha^\dagger)^k(\bar{\alpha}^\dagger)^lp^m\bar{p}_{u_2}\ldots\bar{p}_{u_n}\sum_t h_t\vert V\rangle\\
&-im\sum_r a_rY^{(k)(l)rt_2\ldots t_m(n)}(\bar{\alpha}^\dagger)^k(\alpha^\dagger)^l\bar{p}_{t_2}\ldots\bar{p}_{t_m}p^n\sum_t h_t\vert V\rangle
\end{split}
\end{equation}
%

\paragraph{Terms from $c^{(k)(l)(m)(n-1)}(\alpha^\dagger)^k(\bar{\alpha}^\dagger)^lp^m\bar{p}^{n-1}$ :} These terms commute with everything in the vertex and so just become multiplications. The contributions are
\begin{equation}\label{eq:JMinusHhcalfa}
\begin{split}
&-c^{(k)(l)(m)(n-1)}(\alpha^\dagger)^k(\bar{\alpha}^\dagger)^lp^m\bar{p}^{n-1}\sum_t h_t\vert V\rangle\\
&-c^{(l)(k)(n)(m-1)}(\bar{\alpha}^\dagger)^k(\alpha^\dagger)^l\bar{p}^{m-1}p^n\sum_t h_t\vert V\rangle
\end{split}
\end{equation}
where the second term is the mirror of the first.

\subsection{The D - D differential}\label{subsec:DDDifferential}
With the above noted immediate cancellations and collecting terms from \eqref{eq:CalculationJMinusHpdgamma1}, \eqref{eq:JMinusHMp}, \eqref{eq:JMinusHxh}, \eqref{eq:JMinusHhax} and \eqref{eq:JMinusHhcalfa}, in that order, the $[H,J^-]$ differential can be assembled. It produces two expressions, one from $\Delta^\prime$ and one from the mirror $\bar{\Delta}^\prime$. The oscillator bases $(\alpha^\dagger)^k(\bar{\alpha}^\dagger)^l$ and  $(\bar{\alpha}^\dagger)^k(\alpha^\dagger)^l$ and the vertex will be suppressed and the $r_1,\ldots,r_k$ and $s_1,\ldots,s_l$ labels are not summed over. They pick out a certain direction in Fock space. However, the $t_1,\ldots,t_m$ and $u_1,\ldots,u_n$ labels are still summed over.

\paragraph{Contributions from $\Delta^\prime$ and $\hat{x}$}
\begin{equation}\label{eq:HJ-Differential1}
\begin{split}
i&\sum_r p_r\frac{\partial Y^{(k)(l)(m)(n)}}{{\partial\gamma_r}}p^m\bar{p}^n\\
-i&Y^{r_1\ldots r_k s_1\ldots s_l(m)(n)}\bigg(\frac{p_{r_1}}{\gamma_{r_1}}+\ldots+\frac{p_{r_k}}{\gamma_{r_k}}-\frac{p_{s_1}}{\gamma_{s_1}}-\ldots-\frac{p_{s_l}}{\gamma_{s_l}}\bigg)p^m\bar{p}^n\\
+i&Y^{(k)(l)(m)u_1\ldots u_n}p^m\bar{p}_{u_1}\ldots\bar{p}_{u_n}\bigg(\frac{p_{u_1}}{\gamma_{u_1}}+\ldots+\frac{p_{u_n}}{\gamma_{u_n}}\bigg)\\
-i&n\sum_r a_rY^{(k)(l)(m)ru_2\ldots u_n}p^m\bar{p}_{u_2}\ldots\bar{p}_{u_n}\sum_t h_t\\
-&c^{(k)(l)(m)(n-1)}p^m\bar{p}^{n-1}\sum_t h_t
\end{split}
\end{equation}

\paragraph{Contributions from $\bar{\Delta}^\prime$ and $\hat{\bar x}$}
\begin{equation}\label{eq:HJ-Differential2}
\begin{split}
i&\sum_r p_r\frac{\partial Y^{(k)(l)(m)(n)}}{{\partial\gamma_r}}\bar{p}^m p^n\\
+i&Y^{r_1\ldots r_k s_1\ldots s_l(m)(n)}\bigg(\frac{p_{r_1}}{\gamma_{r_1}}+\ldots+\frac{p_{r_k}}{\gamma_{r_k}}-\frac{p_{s_1}}{\gamma_{s_1}}-\ldots-\frac{p_{s_l}}{\gamma_{s_l}}\bigg)\bar{p}^m p^n\\
+&iY^{(k)(l)t_1\ldots t_m(n)}\bar{p}_{t_1}\ldots\bar{p}_{t_m}\bigg(\frac{p_{t_1}}{\gamma_{t_1}}+\ldots+\frac{p_{t_m}}{\gamma_{t_m}}\bigg)p^n\\
-i&m\sum_r a_rY^{(k)(l)rt_2\ldots t_m(n)}\bar{p}_{t_2}\ldots\bar{p}_{t_m}p^n\sum_t h_t\\
-&c^{(l)(k)(n)(m-1)}\bar{p}^{m-1}p^n\sum_t h_t
\end{split}
\end{equation}
The differential commutator \eqref{eq:DifferentialSideB} is computed in the same way and the result are equations that are complex conjugates of \eqref{eq:HJ-Differential1} and \eqref{eq:HJ-Differential2}. They are completely equivalent.

The cubic differential is particularly interesting, having no source. Next we treat it in detail.

\subsection{The cubic differential recomputed}\label{subsec:CubicDifferentialRecomputed}
For the cubic vertex we use the ans{\"a}tze
\begin{equation}\label{eq:AnsatzDeltaCubic}
\begin{split}
\Delta_3&=Y^{(k)(l)}\left((\alpha^\dagger)^k(\bar{\alpha}^\dagger)^l\mathbb{P}^m\mathbb{\bar{P}}^n+(\bar{\alpha}^\dagger)^k(\alpha^\dagger)^l\mathbb{\bar{P}}^m\mathbb{P}^n\right)\\
\hat{x}_3&=a^r x_r+c^{(k)(l)}(\alpha^\dagger)^{k}(\bar{\alpha}^\dagger)^{l}\mathbb{P}^m\mathbb{\bar{P}}^{n-1}\\
\hat{\bar{x}}_3&=\bar{a}^r \bar{x}_r+\bar{c}^{(k)(l)}(\bar{\alpha}^\dagger)^k(\alpha^\dagger)^l\mathbb{\bar{P}}^m\mathbb{P}^{n-1}
\end{split}
\end{equation}
There is a dependence of the $Y$, $c$ and $\bar c$ on either $m$ or $n$ that is kept implicit. It could be made explicit using a notation $Y^{(k)(l)}_{(m)}$. 

The expression \eqref{eq:HJ-Differential1} is simplified to 
\begin{equation}\label{eq:HJ-Differential1Cubic}
\begin{split}
i&\sum_r p_r\frac{\partial Y^{(k)(l)}}{{\partial\gamma_r}}\mathbb{P}^m\mathbb{\bar{P}}^n+\frac{in}{3}Y^{(k)(l)}\mathbb{P}^m\left(\mathbb{\bar{P}}\sum_r\frac{p_r}{\gamma_r}-\mathbb{P}\sum_r\frac{\bar{p}_r}{\gamma_r}\right)\mathbb{\bar{P}}^{n-1}\\
-i&Y^{r_1\ldots r_k s_1\ldots s_l}\left(\frac{p_{r_1}}{\gamma_{r_1}}+\ldots+\frac{p_{r_k}}{\gamma_{r_k}}-\frac{p_{s_1}}{\gamma_{s_1}}-\ldots-\frac{p_{s_l}}{\gamma_{s_l}}\right)\mathbb{P}^m\mathbb{\bar{P}}^n\\
-\frac{in}{3}&Y^{(k)(l)}\mathbb{P}^m\mathbb{\bar{P}}^{n-1}\sum_rh_r\widetilde{\gamma}_r-c^{(k)(l)}\mathbb{P}^m\mathbb{\bar{P}}^{n-1}\sum_r h_r
\end{split}
\end{equation}
while \eqref{eq:HJ-Differential2} becomes
\begin{equation}\label{eq:HJ-Differential2Cubic}
\begin{split}
i&\sum_r p_r\frac{\partial Y^{(k)(l)}}{{\partial\gamma_r}}\mathbb{\bar{P}}^m\mathbb{P}^n+\frac{im}{3}Y^{(k)(l)}\mathbb{\bar{P}}^{m-1}\left(\mathbb{\bar{P}}\sum_r\frac{p_r}{\gamma_r}-\mathbb{P}\sum_r\frac{\bar{p}_r}{\gamma_r}\right)\mathbb{P}^n\\
+i&Y^{r_1\ldots r_k s_1\ldots s_l}\left(\frac{p_{r_1}}{\gamma_{r_1}}+\ldots+\frac{p_{r_k}}{\gamma_{r_k}}-\frac{p_{s_1}}{\gamma_{s_1}}-\ldots-\frac{p_{s_l}}{\gamma_{s_l}}\right)\mathbb{\bar{P}}^m\mathbb{P}^n\\
-\frac{im}{3}&Y^{(k)(l)}\mathbb{\bar{P}}^{m-1}\mathbb{P}^n\sum_rh_r\widetilde{\gamma}_r-c^{(l)(k)}\mathbb{P}^n\mathbb{\bar{P}}^{m-1}\sum_r h_r
\end{split}
\end{equation}
In both expressions, the line corresponding to line four in \eqref{eq:HJ-Differential1} and \eqref{eq:HJ-Differential2} is identically zero for the cubic vertex. The formulas \eqref{eq:CuReFo1} - \eqref{eq:CuReFo3} of section \ref{subsec:CubicResummationFormulas} are used when computing the $\gamma$-derivatives of $\mathbb{P}$ and $\mathbb{\bar{P}}$. Further simplifications result from using the formulas \eqref{eq:CuReFo4} and \eqref{eq:CuReFo5}.

\begin{equation}\label{eq:HJ-Differential1CubicSimplified}
\begin{split}
i&\sum_r p_r\frac{\partial Y^{(k)(l)}}{{\partial\gamma_r}}\mathbb{P}^m\mathbb{\bar{P}}^n+\frac{2in}{3}Y^{(k)(l)}\mathbb{P}^m\mathbb{\bar{P}}^n\sum_r\frac{p_r}{\gamma_r}\\
-i&Y^{r_1\ldots r_k s_1\ldots s_l}\left(\frac{p_{r_1}}{\gamma_{r_1}}+\ldots+\frac{p_{r_k}}{\gamma_{r_k}}-\frac{p_{s_1}}{\gamma_{s_1}}-\ldots-\frac{p_{s_l}}{\gamma_{s_l}}\right)\mathbb{P}^m\mathbb{\bar{P}}^n+\Gamma^{-1} c^{(k)(l)}\mathbb{P}^{m+1}\mathbb{\bar{P}}^{n}
\end{split}
\end{equation}
and
\begin{equation}\label{eq:HJ-Differential2CubicSimplified}
\begin{split}
i&\sum_r p_r\frac{\partial Y^{(k)(l)}}{{\partial\gamma_r}}\mathbb{\bar{P}}^m\mathbb{P}^n+\frac{2im}{3}Y^{(k)(l)}\mathbb{\bar{P}}^m\mathbb{P}^n\sum_r\frac{p_r}{\gamma_r}\\
+i&Y^{r_1\ldots r_k s_1\ldots s_l}\left(\frac{p_{r_1}}{\gamma_{r_1}}+\ldots+\frac{p_{r_k}}{\gamma_{r_k}}-\frac{p_{s_1}}{\gamma_{s_1}}-\ldots-\frac{p_{s_l}}{\gamma_{s_l}}\right)\mathbb{\bar{P}}^m\mathbb{P}^n+\Gamma^{-1} c^{(l)(k)}\mathbb{P}^{n+1}\mathbb{\bar{P}}^{m}\\
\end{split}
\end{equation}
Since there is no source for the cubic differential, factors of $\mathbb{P}$ and $\mathbb{\bar{P}}$ can now be divided out, but there remains a dependence on $m$ an $n$ through the balance equation $k-l=n-m$. This finally gives

\begin{equation}\label{eq:HJ-Differential1CubicFactoredP}
\begin{split}
i&\sum_r p_r\frac{\partial Y^{(k)(l)}}{{\partial\gamma_r}}+\frac{2in}{3}Y^{(k)(l)}\sum_r\frac{p_r}{\gamma_r}\\
-i&Y^{r_1\ldots r_k s_1\ldots s_l}\left(\frac{p_{r_1}}{\gamma_{r_1}}+\ldots+\frac{p_{r_k}}{\gamma_{r_k}}-\frac{p_{s_1}}{\gamma_{s_1}}-\ldots-\frac{p_{s_l}}{\gamma_{s_l}}\right)+c^{(k)(l)}\frac{\mathbb{P}}{\Gamma}=0
\end{split}
\end{equation}
and
\begin{equation}\label{eq:HJ-Differential2CubicFactoredP}
\begin{split}
i&\sum_r p_r\frac{\partial Y^{(k)(l)}}{{\partial\gamma_r}}+\frac{2im}{3}Y^{(k)(l)}\sum_r\frac{p_r}{\gamma_r}\\
+i&Y^{r_1\ldots r_k s_1\ldots s_l}\left(\frac{p_{r_1}}{\gamma_{r_1}}+\ldots+\frac{p_{r_k}}{\gamma_{r_k}}-\frac{p_{s_1}}{\gamma_{s_1}}-\ldots-\frac{p_{s_l}}{\gamma_{s_l}}\right)+c^{(l)(k)}\frac{\mathbb{P}}{\Gamma}=0
\end{split}
\end{equation}
Both these differential equations must be satisfied for any cubic vertex. We note that they are linear in transverse momentum. Note that $c^{(k)(l)}$ and $c^{(l)(k)}$ are not equal. For vertices with minimal number of transverse derivatives ($m=0$) it turns out that $c^{(l)(k)}=0$. The equations will be solved in the next section.

\section{Systematics of cubic vertices}\label{sec:SystematicsCubicVertices}
The cubic vertex functions can be systematically listed indexed by $(k,l)$ with $k=1,2,3,\ldots$ and $l\leq k$. The powers of transverse momenta $\mathbb{P}^m$ and $\mathbb{\bar{P}}^n$ are determined by the balance equation $n-m=k-l$. On the first level $k=1$ we have
\begin{equation}\label{eq:VertexFunctionListK1}
k=1:\quad l=
\begin{cases}
	0:\quad n-m=1\quad\begin{cases}n\;=1\;2\;3\ldots\\m=0\;1\;2\ldots\end{cases}\\
	1:\quad n-m=0\quad\begin{cases}n\;=0\\m=0\end{cases}\\
\end{cases}
\end{equation}
The minimal terms corresponding to lowest powers of transverse momenta are

\begin{align}\label{eq:VertexFunctionsK1}
\begin{split}
&\begin{cases}
	\Delta^\prime=Y^r\alpha_r^\dagger\mathbb{\bar{P}}\hskip27pt &\text{where}\quad Y^r=1/\gamma_r\\
	\hat{x}=c^r\alpha_r^\dagger\hskip10pt &\text{where}\quad  c^r=\frac{2i}{3}Y^r\widetilde\gamma_r\\
\end{cases}\\
&\begin{cases}
	\Delta^\prime=Y^{rs}\alpha_r^\dagger\bar{\alpha}_s^\dagger\hskip20pt &\text{where}\quad Y^{rs}=\delta_{rs}
\end{cases}
\end{split}
\end{align}
On the next level we have
\begin{equation}\label{eq:VertexFunctionListK2}
k=2:\quad l=
\begin{cases}
	0:\quad n-m=2\quad\begin{cases}n\;=2\;3\;4\ldots\\m=0\;1\;2\ldots\end{cases}\\
	1:\quad n-m=1\quad\begin{cases}n\;=1\;2\;3\ldots\\m=0\;1\;2\ldots\end{cases}\\
	2:\quad n-m=0\quad\begin{cases}n=0\\m=0\end{cases}
\end{cases}
\end{equation}
The minimal terms corresponding to lowest powers of transverse momenta are
\begin{align}\label{eq:VertexFunctionsK2}
\begin{split}
&\begin{cases}
	\Delta^\prime=Y^{r_1r_2}\alpha_{r_1}^\dagger\alpha_{r_2}^\dagger\mathbb{\bar{P}}\mathbb{\bar{P}}\hskip50pt \text{where}\quad Y^{r_1r_2}=1/\gamma_{r_1}\gamma_{r_2}\\
	\hat{x}=c^{r_1r_2}\alpha_{r_1}^\dagger\alpha_{r_2}^\dagger\mathbb{\bar{P}}\hskip67pt \text{where}\quad c^{r_1r_2}=\frac{2i}{3}Y^{r_1r_2}(\widetilde\gamma_{r_1}+\widetilde\gamma_{r_2})\\
\end{cases}\\
&\begin{cases}
	\Delta^\prime=Y^{r_1r_2s_1}\alpha_{r_1}^\dagger\alpha_{r_2}^\dagger\bar{\alpha}_{s_1}^\dagger\mathbb{\bar{P}}\hskip33pt \text{where}\quad Y^{r_1r_2s_1}=\gamma_{s_1}/\gamma_{r_1}\gamma_{r_2}\\
	\hat{x}=c^{r_1r_2s_1}\alpha_{r_1}^\dagger\alpha_{r_2}^\dagger\bar{\alpha}_{s_1}^\dagger\hskip50pt \text{where}\quad c^{r_1r_2s_1}=\frac{2i}{3}Y^{r_1r_2s_1}(\widetilde\gamma_{r_1}+\widetilde\gamma_{r_2}-\widetilde\gamma_{s_1})\\
\end{cases}\\
&\begin{cases}
	\Delta^\prime=Y^{r_1r_2s_1s_2}\alpha_{r_1}^\dagger\alpha_{r_2}^\dagger\bar{\alpha}_{s_1}^\dagger\bar{\alpha}_{s_2}^\dagger\hskip18pt  \text{where}\quad Y^{r_1r_2s_1s_2}=\delta_{r_1s_1}\delta_{r_2s_2}
\end{cases}
\end{split}
\end{align}
These terms are known from previous work done in the 1980's. The general $(k,l\leq k)$ vertex functions with lowest powers in transverse momenta are
\begin{equation}\label{eq:VertexFunctionsKgeneral}
\begin{cases}
	\Delta^\prime=Y^{r_1\ldots r_ks_1\ldots s_l}\alpha_{r_1}^\dagger\ldots\alpha_{r_k}^\dagger\bar{\alpha}_{s_1}^\dagger\ldots\bar{\alpha}_{s_l}^\dagger\mathbb{\bar{P}}^{k-l}\\
	\quad Y^{r_1\ldots r_ks_1\ldots s_l}=\gamma_{s_1}\ldots\gamma_{s_l}/\gamma_{r_1}\ldots\gamma_{r_k}\\
	\hat{x}=c^{r_1\ldots r_ks_1\ldots s_l}\alpha_{r_1}^\dagger\ldots\alpha_{r_k}^\dagger\bar{\alpha}_{s_1}^\dagger\ldots\bar{\alpha}_{s_l}^\dagger\mathbb{\bar{P}}^{k-l-1}\quad\\
	\quad c^{r_1\ldots r_ks_1\ldots s_l}=\frac{2i}{3}Y^{r_1\ldots r_ks_1\ldots s_l}(\widetilde\gamma_{r_1}\ldots+\widetilde\gamma_{r_k}-\widetilde\gamma_{s_1}\ldots-\widetilde\gamma_{s_l})\\
\end{cases}
\end{equation}
The vertex functions follow an easily discernible pattern. 

\subsection{Non-minimal solutions and field redefinitions}\label{subsec:Non-minimalSolutions}
Non-minimal solutions\footnote{The term {\it minimal} should not be confused with ''minimal'' as in ''minimal coupling''. In the present context, {\it minimal} refers to the contributions to $\Delta$ with lowest powers of transverse momenta. I haven't found a better terminology, although perhaps {\it off-shell} would be an alternative as suggested by the note on field redefinitions below.} to the differential equations \eqref{eq:HJ-Differential1CubicFactoredP} and \eqref{eq:HJ-Differential2CubicFactoredP} have factors of $\mathbb P\mathbb{\bar{P}}$. For $k=1,l=0$ we have for instance the $m=1,n=2$ terms
\begin{align}\label{eq:VertexFunctionsK1NonMin1}
\begin{split}
\quad\begin{cases}
	\Delta=Y^r\big(\alpha_r^\dagger\mathbb{\bar{P}}^2\mathbb{P}+\bar\alpha_r^\dagger\mathbb{P}^2\mathbb{\bar{P}}\big)&\text{where}\quad Y^r=(1/\gamma_r)^3\\
	\hat{x}=c^r\big(2\alpha_r^\dagger\mathbb{\bar{P}}\mathbb{P}+\bar\alpha_r^\dagger\mathbb{P}^2\big) &\text{where}\quad  c^r=\frac{2i}{3}Y^r\widetilde\gamma_r\\
\end{cases}
\end{split}
\end{align}
The general formulas of type $m,n=m+1$ on this level are 
\begin{align}\label{eq:VertexFunctionsK1NonMinGen}
\begin{split}
\quad\begin{cases}
	\Delta=Y^r\big(\alpha_r^\dagger\mathbb{\bar{P}}+\bar\alpha_r^\dagger\mathbb{P}\big)(\mathbb{P}\mathbb{\bar{P}})^m&\text{where}\quad Y^r=(1/\gamma_r)^{(1+2m)}\\
	\hat{x}=c^r\big(n\alpha_r^\dagger\mathbb{P}^m\mathbb{\bar{P}}^m+m\bar\alpha_r^\dagger\mathbb{P}^{(m+1)}\mathbb{\bar{P}}^{(m-1)}\big) &\text{where}\quad  c^r=\frac{2i}{3}Y^r\widetilde\gamma_r\\
\end{cases}
\end{split}
\end{align}
The minimal terms are subsumed with $m=0$.\\

\noindent For $k=2,l=1$ we have for instance with $m=1,n=2$
\begin{align}\label{eq:VertexFunctionsK2NonMin1}
\begin{split}
&\begin{cases}
	\Delta=Y^{r_1r_2s_1}\big(\alpha_{r_1}^\dagger\alpha_{r_2}^\dagger\bar{\alpha}_{s_1}^\dagger\mathbb{\bar{P}}^2\mathbb{P}+\bar{\alpha}_{r_1}^\dagger\bar{\alpha}_{r_2}^\dagger\alpha_{s_1}^\dagger\mathbb{P}^2\mathbb{\bar{P}}\big)\\
\quad Y^{r_1r_2s_1}=(\gamma_{s_1}/\gamma_{r_1}\gamma_{r_2})^3\\
	\hat{x}=c^{r_1r_2s_1}\big(2\alpha_{r_1}^\dagger\alpha_{r_2}^\dagger\bar{\alpha}_{s_1}^\dagger\mathbb{\bar{P}}\mathbb{P}+\bar{\alpha}_{r_1}^\dagger\bar{\alpha}_{r_2}^\dagger\alpha_{s_1}^\dagger\mathbb{P}^2\big)\\
\quad c^{r_1r_2s_1}=\frac{2i}{3}Y^{r_1r_2s_1}(\widetilde\gamma_{r_1}+\widetilde\gamma_{r_2}-\widetilde\gamma_{s_1})\\
\end{cases}
\end{split}
\end{align}
The general formulas of type $m,n=m+1$ on this level are 
\begin{align}\label{eq:VertexFunctionsK2Gen}
\begin{split}
&\begin{cases}
	\Delta=Y^{r_1r_2s_1}\big(\alpha_{r_1}^\dagger\alpha_{r_2}^\dagger\bar{\alpha}_{s_1}^\dagger\mathbb{\bar{P}}+\bar{\alpha}_{r_1}^\dagger\bar{\alpha}_{r_2}^\dagger\alpha_{s_1}^\dagger\mathbb{P}\big)(\mathbb{P}\mathbb{\bar{P}})^m\\
\quad Y^{r_1r_2s_1}=(\gamma_{s_1}/\gamma_{r_1}\gamma_{r_2})^{(1+2m)}\\
	\hat{x}=c^{r_1r_2s_1}\big(n\alpha_{r_1}^\dagger\alpha_{r_2}^\dagger\bar{\alpha}_{s_1}^\dagger\mathbb{P}^m\mathbb{\bar{P}}^m+m\bar{\alpha}_{r_1}^\dagger\bar{\alpha}_{r_2}^\dagger\alpha_{s_1}^\dagger\mathbb{P}^{(m+1)}\mathbb{\bar{P}}^{(m-1)}\big)\\
\quad c^{r_1r_2s_1}=\frac{2i}{3}Y^{r_1r_2s_1}(\widetilde\gamma_{r_1}+\widetilde\gamma_{r_2}-\widetilde\gamma_{s_1})\\
\end{cases}
\end{split}
\end{align}
Again, the minimal terms are subsumed with $m=0$.

\subsubsection*{The complete cubic vertex}
Finally, we list the general vertex functions using the abbreviations
\begin{equation}\label{eq:AlphaAbbreviations}
\boldsymbol \alpha_{r_k}^\dagger=\alpha_{r_1}^\dagger\ldots\alpha_{r_k}^\dagger\quad\text{ and }\quad\bar{\boldsymbol \alpha}_{s_l}^\dagger=\bar\alpha_{s_1}^\dagger\ldots\bar\alpha_{s_l}^\dagger
\end{equation}
\begin{align}\label{eq:VertexFunctionsGeneral}
\begin{split}
&\begin{cases}
	\Delta=Y^{r_1\ldots r_ks_1\ldots s_l}\big(\boldsymbol \alpha_{r_k}^\dagger\bar{\boldsymbol \alpha}_{s_l}^\dagger\mathbb{P}^m\mathbb{\bar{P}}^n+\bar{\boldsymbol \alpha}_{r_k}^\dagger\boldsymbol \alpha_{s_l}^\dagger\mathbb{\bar{P}}^m\mathbb{P}^n\big)\\
\quad Y^{r_1\ldots r_ks_1\ldots s_l}=(\gamma_{s_1}\ldots\gamma_{s_l}/\gamma_{r_1}\ldots\gamma_{r_k})^{(\frac{n+m}{n-m})}\quad\text{for}\; n\not=m\\
\quad Y^{r_1\ldots r_ks_1\ldots s_k}=\delta_{r_1s_1}\delta_{r_2s_2}\cdot\ldots\cdot\delta_{r_ks_k}\quad\text{for}\; n=m\;\text{and}\; l=k\\
	\hat{x}=c^{r_1\ldots r_ks_1\ldots s_l}\big(\frac{n}{n-m}\boldsymbol \alpha_{r_k}^\dagger\bar{\boldsymbol \alpha}_{s_l}^\dagger\mathbb{P}^m\mathbb{\bar{P}}^{(n-1)}+\frac{m}{n-m}\bar{\boldsymbol \alpha}_{r_k}^\dagger\boldsymbol \alpha_{s_l}^\dagger\mathbb{\bar{P}}^{(m-1)}\mathbb{P}^{n}\big)\\
\quad c^{r_1\ldots r_ks_1\ldots s_l}=\frac{2i}{3}Y^{r_1\ldots r_ks_1\ldots s_l}(\widetilde\gamma_{r_1}\ldots+\widetilde\gamma_{r_k}-\widetilde\gamma_{s_1}\ldots-\widetilde\gamma_{s_l})\\
\end{cases}
\end{split}
\end{align}
These expressions subsume all the previous given ones. 

The solutions are dimensionally correct since the overall power of $\gamma$ in the $Y$-functions is $(\frac{n+m}{n-m})(l-k)=-(n+m)$. Therefore the $\Delta$ carry overall power of momenta $n+m$, this dimension compensated by the suppressed factor of $\kappa^{n+m}$ mentioned in section \ref{subsec:AnsatzDelta}.

We can now explicate the terms $c^{(k)(l)}$ and $c^{(l)(k)}$ in equations \eqref{eq:HJ-Differential1CubicFactoredP} and \eqref{eq:HJ-Differential2CubicFactoredP} as
\begin{align}
c^{(k)(l)}&=\frac{n}{n-m}c^{r_1\ldots r_ks_1\ldots s_l}\\
c^{(l)(k)}&=\frac{m}{n-m}c^{r_1\ldots r_ks_1\ldots s_l}
\end{align}
where the $c^{(l)(k)}$ terms are zero for minimal powers of transverse momentum ($m=0$).
\subsubsection*{A note on field redefinitions}
Any cubic interaction term containing a factor of $\mathbb P\mathbb{\bar{P}}$ can be removed by a field redefinition. This follows from the identity (see appendix formula \eqref{eq:CuReFo4})
\begin{equation}\label{eq:AppCuReFo4}
\mathbb{P}\mathbb{\bar P}=-\Gamma\sum_{r=1}^3h_r
\end{equation}
Suppose we have a cubic term of the form
\begin{equation}\label{eq:CubicEx}
\mathbb{P}\mathbb{\bar P}\bar\phi_1\phi_2\phi_3
\end{equation}
other momenta not shown. Then using the identity we get
\begin{equation}\label{eq:FieldReDef}
\begin{split}
\mathbb{P}\mathbb{\bar P}\bar\phi_1\phi_2\phi_3&\sim-\Gamma\sum_{r=1}^3h_r\bar\phi_1\phi_2\phi_3\\
&\sim-\Gamma(h_1\bar\phi_1\phi_2\phi_3+\bar\phi_1h_2\phi_2\phi_3+\bar\phi_1\phi_2h_3\phi_3)
\end{split}
\end{equation}
The field equations are $h\bar\phi={\mathcal O}(\bar\phi\phi)$ and $h\phi={\mathcal O}(\phi\phi)$, so the terms are zero on-shell to cubic order. From another perspective, making field redefinitions of the form $\phi\rightarrow\phi+F(\phi,\phi)$ and $\bar\phi\rightarrow\bar\phi+F(\phi,\bar\phi)$, the terms can be cancelled against variations of the kinetic term $\phi h\bar\phi$. This can be made more exact by introducing a field redefinition cubic vertex $\vert F_{123}\rangle$. Such a vertex must satisfy all kinematic constraints. This fixes the momentum structure except for the detailed dependence of the $\gamma_r$. It is this freedom in $\gamma$ structure that allows for transforming away any tentative interactions containing products of $\mathbb P\mathbb{\bar{P}}$. Therefore, at least to cubic order, it is enough to consider minimal vertex functions.
\subsection{Solution of the cubic differential}\label{subsec:SolutionCubicDifferential}
The solutions listed in the previous section can be proved to be correct in the following way. First apply the re-summation formula to the derivative term to obtain
\begin{equation}\label{eq:ResummedDerivativeY}
\sum_r p_r\frac{\partial Y^{(k)(l)}}{{\partial\gamma_r}}=\frac{1}{3}\sum_r \gamma_r\frac{\partial Y^{(k)(l)}}{{\partial\gamma_r}}\sum_{s=1}^3\frac{p_s}{\gamma_s}+\frac{1}{3}\sum_r \gamma_r\widetilde{\gamma}_r\frac{\partial Y^{(k)(l)}}{{\partial\gamma_r}}\frac{\mathbb{P}}{\Gamma}
\end{equation}
The two terms on the right hand side are independent in the sense that there is no way to rewrite the one in terms of the other. One can think of the two objects
\begin{equation}\label{eq:IndependentBasis}
\sum_{s=1}^3\frac{p_s}{\gamma_s}\quad\text{ and }\quad\frac{\mathbb{P}}{\Gamma}
\end{equation}
as an independent basis in the two-dimensional space spanned by the constrained momenta. Comparing with the differential equations \eqref{eq:HJ-Differential1CubicFactoredP} and \eqref{eq:HJ-Differential2CubicFactoredP} suggests solving for the non-diagonal term. We get
\begin{equation}\label{eq:NonDiagonalYterm}
\begin{split}
i&Y^{r_1\ldots r_k s_1\ldots s_l}\left(\frac{p_{r_1}}{\gamma_{r_1}}+\ldots+\frac{p_{r_k}}{\gamma_{r_k}}-\frac{p_{s_1}}{\gamma_{s_1}}-\ldots-\frac{p_{s_l}}{\gamma_{s_l}}\right)\\
=&\frac{i(n-m)}{3}Y^{(k)(l)}\sum_r\frac{p_r}{\gamma_r}+\frac{1}{2}\Big(c^{(k)(l)}-c^{(l)(k)}\Big)\frac{\mathbb{P}}{\Gamma}
\end{split}
\end{equation}
Inserting this formula back into either of the differential equations \eqref{eq:HJ-Differential1CubicFactoredP} or \eqref{eq:HJ-Differential2CubicFactoredP} and using the re-summed derivative \eqref{eq:ResummedDerivativeY} and the independence of \eqref{eq:IndependentBasis}, we get the two differential equations
\begin{subequations}\label{eq:ReducedDifferentialEqs}
\begin{equation}\label{eq:ReducedDifferentialEqsA}
\sum_r \gamma_r\frac{\partial Y^{(k)(l)}}{{\partial\gamma_r}}=-(m+n)Y^{(k)(l)}
\end{equation}
\begin{equation}\label{eq:ReducedDifferentialEqsA}
\sum_r \gamma_r\widetilde{\gamma_r}\frac{\partial Y^{(k)(l)}}{{\partial\gamma_r}}=\frac{3i}{2}\Big(c^{(k)(l)}+c^{(l)(k)}\Big)
\end{equation}
\end{subequations}
This is a surprising result. The first equation is precisely the same equation that follows from the $j^{+-}$ homogeneity constraint since using \eqref{eq:CuReFo8}
\begin{equation}
\sum_{r=1}^3\gamma_r\frac{\partial}{\partial\gamma_r}\mathbb{P}^m\bar{\mathbb{P}}^n=(m+n)\mathbb{P}^m\bar{\mathbb{P}}^n
\end{equation}
Once the vertex functions $Y^{(k)(l)}$ are determined, the second equation yields the $J^-$ pre-factor functions $c^{(k)(l)}$. 

This shows, contrary to what one would naively have expected, that the $[\mathcal D,\mathcal D]$ part of the cubic Poincar{\'e} algebra, provide no more restrictions than come from the $[\mathcal K,\mathcal D]$ commutators. Or to be precise: The cubic vertex functions are determined by the kinematics. This fact has not been apparent from previous light-front treatments of cubic interactions.

It's a matter of straightforward calculation to show that the solutions given in \eqref{eq:VertexFunctionsGeneral} solves the differential equations \eqref{eq:ReducedDifferentialEqs}.

\subsection{The rest of the cubic algebra}\label{subsec:RestCubicAlgebra}
The second expression \eqref{eq:DifferentialSideB} is the complex conjugate of the first \eqref{eq:DifferentialSideA}. For the cubic vertex, which has no source, both differentials should be zero and the second yields no further information. The third expression \eqref{eq:DifferentialSideC} can be simplified provided the first two differentials are zero. The calculation runs as follows.
\begin{equation*}
\begin{split}
&\sum_{r=1}^3\left(\bar{j}_r^-\hat{x}-j_r^-\hat{\bar{x}}\right)\vert V\rangle\\
=&\sum_{r=1}^3\left([\bar{j}_r^-,\hat{x}]-[j_r^-,\hat{\bar{x}}]\right)\vert V\rangle+\sum_{r=1}^3\left(\hat{x}\bar{j}_r^--\hat{\bar{x}}j_r^-\right)\vert V\rangle\\
=&\sum_{r=1}^3\left([\bar{j}_r^-,\hat{x}]-[j_r^-,\hat{\bar{x}}]\right)\vert V\rangle+\sum_{r=1}^3\left(\hat{x}h_r\hat{\bar{x}}-\hat{\bar{x}}h_r\hat{x}\right)\vert V\rangle\\
=&\sum_{r=1}^3\left([\bar{j}_r^-,\hat{x}]-[j_r^-,\hat{\bar{x}}]\right)\vert V\rangle+\sum_{r=1}^3\left(\hat{x}[h_r,\hat{\bar{x}}]-\hat{\bar{x}}[h_r,\hat{x}]\right)\vert V\rangle
\end{split}
\end{equation*}
In the last equality we have used $[\hat{x}_{(\nu)},\hat{\bar{x}}_{(\nu)}]=0$. The right hand side should be read as anything that survives the commutators should act on the vertex. Showing that this expression is zero for the cubic vertex involves a calculational subtlety.

In order to reduce clutter and highlight the relevant details the notation will be simplified.\footnote{By freely removing and inserting for instance $\vert V\rangle$ as need arise.} The prefactors $\hat{x}$ and $\hat{\bar{x}}$ are given in \eqref{eq:AnsatzDeltaCubic}. The commutators are computed for the terms containing $a^s$ and $c^{(k)(l)}$ and $\bar{c}^{(k)(l)}$ separately as follows. In the formulas, factors of $\mathbb{P}$ and $\mathbb{\bar{P}}$ have been factored out.

\paragraph{Terms from $[\bar{j}_r^-,a^sx_s]-[j_r^-,a^s\bar{x}_s]$ :}This results in
\begin{align*}
&-\frac{i}{3}\sum_r^3\frac{1}{\gamma_r}(p_r\bar{x}_r-\bar{p}_rx_r+2M_r)\vert V\rangle\\
&\rightarrow\frac{n}{9}Y^{(k)(l)}\mathbb{P}\sum_r^3\frac{\widetilde\gamma_r\bar{p}_r}{\gamma_r}-\frac{m}{9}Y^{(k)(l)}\mathbb{\bar{P}}\sum_r^3\frac{\widetilde\gamma_rp_r}{\gamma_r}\\
&+\frac{2}{3}Y^{r_1\ldots r_k s_1\ldots s_l}\left(\frac{1}{\gamma_{r_1}}+\ldots+\frac{1}{\gamma_{r_k}}-\frac{1}{\gamma_{s_1}}-\ldots-\frac{1}{\gamma_{s_l}}\right)\mathbb{P}\mathbb{\bar{P}}
\end{align*}
The oscillator basis $(\alpha^\dagger)^k(\bar{\alpha}^\dagger)^l$ is suppressed. Note that there are really two sets of terms here, one with oscillator basis $(\alpha^\dagger)^k(\bar{\alpha}^\dagger)^l$, the other with oscillator basis $(\bar{\alpha}^\dagger)^k(\alpha^\dagger)^l$ corresponding to $\Delta^\prime$ and $\bar\Delta^\prime$.

\paragraph{Terms from $[\bar{j}_r^-,c^{(k)(l)}]-[j_r^-,\bar{c}^{(k)(l)}]$ :}This results in
\begin{align*}
&i\mathbb{P}\sum_r^3\bar{p}_r\frac{\partial c^{(k)(l)}}{\partial\gamma_r}+\frac{2im}{3}c^{(k)(l)}\mathbb{P}\sum_r^3\frac{\bar{p}_r}{\gamma_r}\\
+&ic^{r_1\ldots r_k s_1\ldots s_l}\left(\frac{\bar{p}_{r_1}}{\gamma_{r_1}}+\ldots+\frac{\bar{p}_{r_k}}{\gamma_{r_k}}-\frac{\bar{p}_{s_1}}{\gamma_{s_1}}-\ldots-\frac{\bar{p}_{s_l}}{\gamma_{s_l}}\right)\mathbb{P}\\
&-i\mathbb{\bar{P}}\sum_r^3p_r\frac{\partial\bar{c}^{(k)(l)}}{\partial\gamma_r}-\frac{2in}{3}\bar{c}^{(k)(l)}\mathbb{\bar{P}}\sum_r^3\frac{p_r}{\gamma_r}\\
+&i\bar{c}^{r_1\ldots r_k s_1\ldots s_l}\left(\frac{p_{r_1}}{\gamma_{r_1}}+\ldots+\frac{p_{r_k}}{\gamma_{r_k}}-\frac{p_{s_1}}{\gamma_{s_1}}-\ldots-\frac{p_{s_l}}{\gamma_{s_l}}\right)\mathbb{\bar{P}}
\end{align*}
%

\paragraph{Terms from $\hat{x}[h_r,\hat{\bar{x}}]-\hat{\bar{x}}[h_r,\hat{x}]$ :}This is where the calculational subtlety resides. In \cite{BBL1987} the commutators $\sum_r[h_r,\hat{\bar{x}}]$ and $[h_r,\hat{x}]$ were erroneously taken to be zero. One might be lead to that conclusion if $\sum_rh_r$ is replaced by $-\mathbb{P}\mathbb{{\bar P}}/\Gamma$. But that cannot be done inside a commutator since $x$ and $\bar{x}$ differentiates the transverse derivatives and these are not independent due to momentum conservation. Instead
\begin{equation}
\sum_r^3[h_r,\hat{\bar{x}}]=-\frac{i}{3}\sum_r^3\frac{\bar{p}_r}{\gamma_r}\quad\text{ and }\quad\sum_r^3[h_r,\hat{x}]=-\frac{i}{3}\sum_r^3\frac{p_r}{\gamma_r}
\end{equation}
Then the $a$-terms from the multiplying prefactors yield zero. This comes about since they commute through $e^\Delta$ and the net result is zero due to the zero orbital angular momentum of the vacuum.
\begin{align*}
\Big(\sum_s^3a^sx_s\sum_r^3[h_r,\hat{\bar{x}}]-\sum_s^3a^s\bar{x}_s\sum_r^3[h_r,\hat{x}]\Big)e^\Delta\vert\varnothing\rangle\sim e^\Delta\sum_r^3(x_r\bar{p}_r-\bar{x}_rp_r)\vert\varnothing\rangle=0
\end{align*}
However, from the $c$-terms there result
\begin{equation}
-\frac{1}{3}c^{(k)(l)}\mathbb{P}\sum_r^3\frac{\bar{p}_r}{\gamma_r}+\frac{1}{3}\bar{c}^{(k)(l)}\mathbb{\bar{P}}\sum_r^3\frac{p_r}{\gamma_r}
\end{equation}
Putting it all together the equation to satisfy is
\begin{equation}\label{eq:DynLorentzCommutators}
\begin{split}
&i\mathbb{P}\sum_r^3\bar{p}_r\frac{\partial c^{(k)(l)}}{\partial\gamma_r}+i\frac{2m-1}{3}c^{(k)(l)}\mathbb{P}\sum_r^3\frac{\bar{p}_r}{\gamma_r}+\frac{n}{9}\mathbb{P}Y^{(k)(l)}\sum_r^3\frac{\widetilde\gamma_r\bar{p}_r}{\gamma_r}\\
+&ic^{r_1\ldots r_k s_1\ldots s_l}\left(\frac{\bar{p}_{r_1}}{\gamma_{r_1}}+\ldots+\frac{\bar{p}_{r_k}}{\gamma_{r_k}}-\frac{\bar{p}_{s_1}}{\gamma_{s_1}}-\ldots-\frac{\bar{p}_{s_l}}{\gamma_{s_l}}\right)\mathbb{P}\\
-&i\mathbb{\bar{P}}\sum_r^3p_r\frac{\partial\bar{c}^{(k)(l)}}{\partial\gamma_r}-i\frac{2n-1}{3}\bar{c}^{(k)(l)}\mathbb{\bar{P}}\sum_r^3\frac{p_r}{\gamma_r}-\frac{m}{9}Y^{(k)(l)}\mathbb{\bar{P}}\sum_r^3\frac{\widetilde\gamma_rp_r}{\gamma_r}\\
+&i\bar{c}^{r_1\ldots r_k s_1\ldots s_l}\left(\frac{p_{r_1}}{\gamma_{r_1}}+\ldots+\frac{p_{r_k}}{\gamma_{r_k}}-\frac{p_{s_1}}{\gamma_{s_1}}-\ldots-\frac{p_{s_l}}{\gamma_{s_l}}\right)\mathbb{\bar{P}}\\
+&\frac{2}{3}Y^{r_1\ldots r_k s_1\ldots s_l}\left(\frac{1}{\gamma_{r_1}}+\ldots+\frac{1}{\gamma_{r_k}}-\frac{1}{\gamma_{s_1}}-\ldots-\frac{1}{\gamma_{s_l}}\right)\mathbb{P}\mathbb{\bar{P}}=0
\end{split}
\end{equation}
It can be checked that the solutions given above satisfy these equations. This finally shows cubic Poincar{\'e} invariance.
\subsection{Open questions regarding the cubics}\label{subsec:OpenQuestionsCubic}
There is a huge redundancy in the cubic vertex functions as presented here. In string theory, which only involves quadratic forms in oscillators and momenta, the exponential of the vertex functions is needed in order to generate higher spin (massive) interactions. In the present case, the exponential is not really needed because higher spin interactions are generated from the higher order polynomial form vertex functions. Still, in computing the commutators it is convenient that derivatives of exponentials produce a factor times the same exponential. It is clear that the non-linear Poincar{\'e} algebra is not very restrictive on the cubic level. The vertices that we have listed here generates all possible cubic vertices between Fock field components in $\vert\Theta\rangle$. Among these there will be terms that can be removed by field redefinitions.

\section{Summary of conventions and formulas}\label{sec:ConventionsFormulas}
\subsection{Light-front coordinates and derivatives}\label{subsec:LFCoordinatesDerivatives}
Coordinates are give by
\begin{equation}\label{eq:LightFrontCoordinates}
\begin{split}
x^+&=\frac{1}{\sqrt 2}(x^0+x^3)\hspace{25pt} x^-=\frac{1}{\sqrt 2}(x^0-x^3)\\
x&=\frac{1}{\sqrt 2}(x^1+ix^2)\hspace{28pt} \bar{x}=\frac{1}{\sqrt 2}(x^1-ix^2)
\end{split}
\end{equation}
Derivatives are given by
\begin{equation}\label{eq:LightFrontDerivatives}
\begin{split}
\partial_+&=\frac{\partial}{\partial x^+}=\frac{1}{\sqrt 2}(\partial_0+\partial_3)=-\partial^-\hspace{23pt} \partial_-=\frac{\partial}{\partial x^-}=\frac{1}{\sqrt 2}(\partial_0-\partial_3)=-\partial^+\\
\partial&=\frac{\partial}{\partial \bar{x}}=\frac{1}{\sqrt 2}(\partial_1+i\partial_2)\hspace{69pt} \bar{\partial}=\frac{\partial}{\partial x}=\frac{1}{\sqrt 2}(\partial_1-i\partial_2)
\end{split}
\end{equation}
With a mostly-plus Minkowski metric $-+++$, the light-front scalar product becomes
\begin{equation}\label{eq:LightFrontScalarProduct}
\begin{split}
A_\mu B^\mu&=A\bar{B}+\bar{A} B+A_-B^-+A_+B^+\\
&=A\bar{B}+\bar{A} B-A^+B^--A^-B^+
\end{split}
\end{equation}
in particular
\begin{equation}\label{eq:LFBox}
\square=\partial_\mu\partial^\mu=2(-\partial_+\partial_-+\partial\bar\partial)
\end{equation}
The transverse oscillators are
\begin{equation}\label{eq:LightFrontOscillators}
\begin{split}
\alpha&=\frac{1}{\sqrt 2}(\alpha_1+i\alpha_2)\hspace{28pt} \bar{\alpha}=\frac{1}{\sqrt 2}(\alpha_1-i\alpha_2)\\
\alpha^\dagger&=\frac{1}{\sqrt 2}(\alpha_1^\dagger+i\alpha_2^\dagger)\hspace{28pt} \bar{\alpha}^\dagger=\frac{1}{\sqrt 2}(\alpha_1^\dagger-i\alpha_2^\dagger)
\end{split}
\end{equation}
To avoid confusion, note that the dagger $\dagger$ has no operational significance here, it just denotes a creation operator. Thus $\alpha^\dagger$ is not the hermitean conjugate of $\alpha$.

\subsection{Connection to amplitude notation}\label{subsec:AmplitudeNotation}
The transverse momentum constructions $\mathbb{P}$ and $\bar{\mathbb{P}}$ are closely related to the spinors $\lambda_a$ and $\tilde\lambda_{\dot{a}}$ used in MHV amplitude research. Using the Pauli matrices $\sigma^\mu=(\mathbf1,\sigma^i)$ we have
\begin{equation}\label{eq:SpinorConversion1}
p_{a\dot{a}}=p_\mu\sigma^\mu_{a\dot{a}}=\begin{pmatrix}p_0+p_3&p_1-ip_2\cr p_1+ip_2&p_0-p_3\end{pmatrix}=\sqrt2\begin{pmatrix}-p^-&\bar p\cr p&-p^+\end{pmatrix}
\end{equation}
On-shell we get
\begin{equation}\label{eq:SpinorConversion2}
\sqrt2\begin{pmatrix}-p\bar p/\gamma&\bar p\cr p&-\gamma\end{pmatrix}
\end{equation}
which can be factored into the holomorphic and anti-holomorphic spinors
\begin{equation}\label{eq:SpinorConversion3}
|p]_a=\frac{\sqrt[4]2}{\sqrt\gamma}\begin{pmatrix}\bar p\\-\gamma\end{pmatrix}\quad\text{and}\quad\langle p|_{\dot a}=\frac{\sqrt[4]2}{\sqrt\gamma}\begin{pmatrix}p\,,&-\gamma\end{pmatrix}
\end{equation}
Then contractin with the anti-symmetric tensors $\epsilon^{ab}$ and $\epsilon^{\dot{a}\dot{b}}$ we get
\begin{align}\label{eq:SpinorConversion4}
\langle p_1 p_2\rangle&=-\frac{\sqrt{2}}{\sqrt{\gamma_1\gamma_2}}\big(p_1\gamma_2-p_2\gamma_1\big)\\
[p_1 p_2]&=\frac{\sqrt{2}}{\sqrt{\gamma_1\gamma_2}}\big(\bar p_1\gamma_2-\bar p_2\gamma_1\big)
\end{align}
Using momentum conservation $p_1\gamma_2-p_2\gamma_1=\mathbb{P}$ and $\bar{p}_1\gamma_2-\bar{p}_2\gamma_1=\mathbb{\bar{P}}$.
\subsection{Shorthand notation}\label{subsec:ShorthandNotation}
Convenient shorthand
\begin{equation}\label{eq:ShorthandNotation}
\begin{split}
d^2x&=dxd\bar{x} \hspace{125pt}d^2p=dpd\bar{p}\\
d^3x&=dx^-d^2x\hspace{114pt} d^3p=dp^+d^2p\\
\delta^2(x-y)&=\delta(x-y)\delta(\bar{x}-\bar{y})\hspace{44pt} \delta^2(p+q)=\delta(p+q)\delta(\bar{p}+\bar{q})\\
\delta^3(x-y)&=\delta(x^--y^-)\delta^2(x-y)\hspace{25pt} \delta^3(p+q)=\delta(p^++q^+)\delta^2(p+q)
\end{split}
\end{equation}

\subsection{Fields and Fourier transforms}\label{subsec:FieldsFourierTransforms}

Consider two real fields $\phi_k(x)$ such as the transverse components of a light-front higher helicity field. The basic equal $x^+$ commutator is
\begin{equation}\label{eq:BasicEqualTimeCommutator}
[\phi_k(x),\partial_y^+\phi_l(y)]_{x^+=y^+}=i\delta^3(x-y)\delta_{kl}
\end{equation}
The Fourier transform pairs are
\begin{equation}\label{eq:FourierTransformPairs}
\begin{split}
\phi_k(x)&=\frac{1}{(2\pi)^{3/2}}\int d^3p\phi_k(p)e^{ip\cdot x}\\
\phi_k(p)&=\frac{1}{(2\pi)^{3/2}}\int d^3x\phi_k(x)e^{-ip\cdot x}
\end{split}
\end{equation}
A complex field is introduced by defining
\begin{equation}\label{eq:ComplexFieldDefinition}
\phi=\frac{1}{\sqrt2}(\phi_1+i\phi_2)\quad\quad\quad\bar{\phi}=\frac{1}{\sqrt2}(\phi_1-i\phi_2)
\end{equation}
both for $x$\,-\,space and $p$\,-\,space fields. In terms of the complex field, the non-zero equal time commutator become
\begin{equation}\label{eq:XComplexEqualTimeCommutator}
[\phi(x),\partial_y^+\bar{\phi}(y)]_{x^+=y^+}=i\delta^3(x-y)
\end{equation}
It can be derived directly by Dirac analysis of the light-front Lagrangian density
\begin{equation*}
\mathcal{L}=\frac{1}{2}\big(\partial_+\phi\partial_-\bar{\phi}+\partial_+\bar{\phi}\partial_-\phi\big)-\partial\phi\bar{\partial}\bar{\phi}
\end{equation*}
and subsequent quantization. In the process one finds the Hamiltonian
\begin{equation*}
H=\int dx^-d^2x\partial\phi\bar{\partial}\bar{\phi}
\end{equation*}
It is a peculiarity of light-front field theory that the conjugate momenta $\pi=\partial_-\phi$ and $\bar\pi=\partial_-\bar\phi$ don't appear explicitly in the Hamiltonian.

Corresponding to \eqref{eq:XComplexEqualTimeCommutator} we get the $p$\,-\,space equal time commutator (note that $i\partial_y^+e^{-iq\cdot y}=q^+e^{-iq\cdot y}$)
\begin{equation}\label{eq:PEqualTimeCommutatorDerivation}
\begin{split}
[\phi_k(p),q^+\phi_l(q)]&=\frac{1}{(2\pi)^3}\int d^3xd^3y[\phi_k(x),i\partial_y^+\phi_l(y)]e^{-i(p\cdot x+q\cdot y)}\\
&=-\frac{1}{(2\pi)^3}\int d^3xd^3y\delta_{kl}\delta^3(x-y)e^{-i(p\cdot x+q\cdot y)}\\
&=-\delta^3(p+q)
\end{split}
\end{equation}
For complex fields this yields the convenient form
\begin{equation}\label{eq:PComplexEqualTimeCommutator}
[\phi(p),\bar{\phi}(q)]_{x^+=y^+}=-\frac{\delta^3(p+q)}{q^+}
\end{equation}

\subsection{The Fock space}\label{subsec:FockSpace}
The vacuum is really a double vacuum so we write $\vert 0\rangle=\vert0,\overline{0}\rangle$. The excited states are given by (the bar here distinguish the states excited by $\bar\alpha^\dagger$ from the states excited by $\alpha^\dagger$)
\begin{equation}\label{eq:FockStates}
\begin{split}
\alpha^\dagger\big\vert n,\overline{m}\rangle&=\sqrt{n+1}\big\vert n+1,\overline{m}\rangle\quad\quad \bar\alpha^\dagger\big\vert n,\overline{m}\rangle=\sqrt{m+1}\big\vert n,\overline{m+1}\rangle\\
\bar\alpha\big\vert n,\overline{m}\rangle&=\sqrt{n}\big\vert n,\overline{m-1}\rangle\quad\quad\quad\quad \alpha\big\vert n,\bar m\rangle=\sqrt{m}\big\vert n-1,\overline{m}\rangle
\end{split}
\end{equation}
With $M$ and $N$ defined in \eqref{eq:G0Algebra} we have
\begin{equation}
\begin{split}
N\big\vert n,\overline{m}\rangle&=(n+m+1)\big\vert n,\overline{m}\rangle\\
M\big\vert n,\overline{m}\rangle&=(n-m)\big\vert n,\overline{m}\rangle
\end{split}
\end{equation}
The states with constant $N$ eigenvalue form a module for the algebra $g^0=\mathrm{U}(2)$. They can also be parameterized as $\mathrm{SO}(3)$ angular momentum states $\vert j,m_j\rangle$ with $j=n+m+1$ and $m_j=n-m$.

The full Fock space can be pictured as in Table 1.

\begin{table}[h]\label{fig:StateDiagram}
\centering
\begin{tabular}{c c c c c c c}
  &  &  &  &  &  & $\cdots$ \cr
  &  &  &  &  & $|05\rangle$&\cr
  &  &  &  & $|04\rangle$&\cr
  &  &  & $|03\rangle$&  & $|14\rangle$\cr
  &  & $|02\rangle$&  & $|13\rangle$&\cr
  & $|01\rangle$&  &$|12\rangle$ & &  $|23\rangle$\cr
 $|00\rangle$&  & $|11\rangle$&  & $|22\rangle$&\cr
  & $|10\rangle$&  & $|21\rangle$ & &  $|32\rangle$\cr
  &  & $|20\rangle$&  & $|31\rangle$&\cr
  &  &  & $|30\rangle$&  & $|41\rangle$\cr
  &  &  &  & $|40\rangle$&\cr
  &  &  &  &  & $|50\rangle$\cr
 &  &  &  &  &  & $\cdots$ \cr
\end{tabular}
\caption{The Fock space}
\label{tab:}
\end{table}
\noindent It should be clear that $\mathrm{SO}(3)$ is finitely represented in each column. It is convenient to call the eigenvalue of $N-1$, that is $n+\bar m$, the {\it spin} as it does indeed correspond to the spin of a massive representation. The non-compact $\mathrm{SU}(1,1)$ is infinitely represented in each row. The physical pure helicity states $\vert0\,\lambda\rangle$ and $\vert \lambda\, 0\rangle$ are defined by $T\big\vert n,\bar m\rangle=0$ and these all lie on the border of the diagram. The operator $T^\dagger$ excites the pure helicity states into states with the same helicity but belonging to a different $\mathrm{SO}(3)$ module increasing the spin by $2$.

Furthermore, the set of generators $g^{-1}\oplus g^{0}\oplus g^{+1}$ are reducibly represented on the full Fock space. The two irreducible modules correspond to all odd and all even spin respectively. Adding in the oscillators themselves, it is possible to transform between even and odd spin. The oscillators together with $g^{-1}\oplus g^{0}\oplus g^{+1}$ span the $\mathrm{Osp}(1|4)$ algebra which is consequently represented irreducibly on the full Fock space. Alternatively, the full Fock space can be parameterized in terms of Fronsdal Di and Rac states by reinterpreting $\lambda\rightarrow\lambda/2$. 

A recent thorough discussion of these algebras of relevance to the present work can be found in \cite{BekaertTraubenbergValenzuela2009}.

\subsection{Construction of the delta vertices}\label{subsec:ConstructionDeltaVertices}
In constructing the delta vertices we follow \cite{GrossJevicki1987I}. The delta vertices should convert between different bra and ket Fock spaces. Consider the operator 
\begin{equation}\label{eq:DeltaVertices1}
\vert I_{1^\dagger2^\dagger}\rangle=\exp(\alpha_1^\dagger{\bar\alpha}_2^\dagger+\alpha_2^\dagger{\bar\alpha}_1^\dagger)\vert0,\bar0\rangle_1\vert0,\bar0\rangle_2
\end{equation}
and let it act on the bra state $\langle 0,\bar0\vert_1(\bar\alpha_1)^n\phi$
\begin{align}\label{eq:DeltaVertices2}
\langle 0,\bar0\vert_1(\bar\alpha_1)^n\phi\vert I_{12}\rangle&=\phi\langle 0,\bar0\vert_1(\bar\alpha_1)^n\frac{1}{n!}(\alpha_1^\dagger{\bar\alpha}_2^\dagger)^n\vert0,\bar0\rangle_1\vert0,\bar0\rangle_2\\
&=\phi\langle 0,\bar0\vert_1({\bar\alpha}_2^\dagger)^n\vert0,\bar0\rangle_1\vert0,\bar0\rangle_2=\phi({\bar\alpha}_2^\dagger)^n\vert0,\bar0\rangle_2
\end{align}
which is converted to the ket state $\phi({\bar\alpha}_2^\dagger)^n\vert0,\bar0\rangle_2$. The other delta operators $I_{1^\dagger2}$, $I_{12^\dagger}$ and $I_{12}$  are defined similarly and work analogously.

Next consider the commutator $[\Phi^\dagger(p_1),\Phi(p_2)]$ between two Fock fields 
\begin{align}\label{eq:FockSpaceFields}
\Phi^\dagger(p_1)&=\sum_{\lambda=0}^\infty\frac{1}{\sqrt{\lambda!}}\left(\phi_{\lambda}(p_1)({\bar\alpha}_1^\dagger)^\lambda+\bar{\phi}_{\lambda}(p_1)(\alpha_1^\dagger)^\lambda\right)\\
\Phi(p_2)&=\sum_{\lambda^\prime=0}^\infty\frac{1}{\sqrt{\lambda!}}\left(\phi_{\lambda^\prime}(p_2){\bar\alpha}_2^{\lambda^\prime}+\bar{\phi}_{\lambda^\prime}(p_2)\alpha_2^{\lambda^\prime}\right)
\end{align}
In calculating the commutator, note that the oscillators commute as they refer to different Fock spaces, and that the only non-zero field commutators are between un-bared and bared fields. This yields
\begin{equation}\label{eq:FockSpaceFieldCommutator}
\begin{split}
&[\Phi^\dagger(p_1),\Phi(p_2)]\\
=&\sum_{\lambda=0}^\infty\sum_{\lambda^\prime=0}^\infty\frac{1}{\sqrt{\lambda!{\lambda^\prime}!}}\Big[\phi_{\lambda}(p_1)({\bar\alpha}_1^\dagger)^\lambda+\bar{\phi}_{\lambda}(p_1)(\alpha_1^\dagger)^\lambda,\phi_{\lambda^\prime}(p_2){\bar\alpha}_2^{\lambda^\prime}+\bar{\phi}_{\lambda^\prime}(p_2)\alpha_2^{\lambda^\prime}\Big]\\
=&\sum_{\lambda=0}^\infty\sum_{\lambda^\prime=0}^\infty\frac{1}{\sqrt{\lambda!{\lambda^\prime}!}}\Big(\Big[\phi_{\lambda}(p_1),\bar{\phi}_{\lambda^\prime}(p_2)\Big]({\bar\alpha}_1^\dagger)^\lambda\alpha_2^{\lambda^\prime}+\Big[\bar{\phi}_{\lambda}(p_1),\phi_{\lambda^\prime}(p_2)\Big](\alpha_1^\dagger)^\lambda{\bar\alpha}_2^{\lambda^\prime}\Big)\\
=&\sum_{\lambda=0}^\infty\sum_{\lambda^\prime=0}^\infty\frac{1}{\sqrt{\lambda!{\lambda^\prime}!}}\Big(-\frac{\delta^3(p_1+p_2)}{p_2^+}\delta_{\lambda\lambda^\prime}({\bar\alpha}_1^\dagger)^\lambda\alpha_2^{\lambda^\prime}+\frac{\delta^3(p_2+p_1)}{p_1^+}\delta_{\lambda\lambda^\prime}(\alpha_1^\dagger)^\lambda{\bar\alpha}_2^{\lambda^\prime}\Big)\\
\\
=&-\frac{\delta^3(p_1+p_2)}{p_2^+}\sum_{\lambda=0}^\infty\frac{1}{\lambda!}\Big(({\bar\alpha}_1^\dagger)^\lambda\alpha_2^{\lambda}+(\alpha_1^\dagger)^\lambda{\bar\alpha}_2^{\lambda}\Big)=-\frac{\delta^3(p_1+p_2)}{p_2^+}I_{1^\dagger2}
\end{split}
\end{equation}
Thus the delta operators $I_{1^\dagger2}$ et cetera work as expected.
\subsection{Computation of commutators}\label{subsec:ComputationCommutators}
Here we explain the translation of the Poincar{\'e} commutators of generators to the  computational form used in the paper. Let $k$ be a kinematical generator and $d$ a dynamical generator with corresponding operators $\hat k$ and $\hat d$. Then we have on a certain field $|\Phi_3\rangle$
\begin{equation}\label{eq:ComputationCommutators1}
\begin{split}
\delta_k|\Phi_3\rangle&=\hat k_3|\Phi_3\rangle\\
\delta_d|\Phi_3\rangle&=\hat d_3|\Phi_3\rangle+\langle\Phi_1|\langle\Phi_2|\hat d_{123}|V_{123}\rangle\equiv\langle\Phi_1|\langle\Phi_2|D_{123}\rangle
\end{split}
\end{equation}
Then consider the commutator
\begin{equation*}
[\delta_k,\delta_d]|\Phi_3\rangle=\delta_g|\Phi_3\rangle
\end{equation*}
where $g$ may be zero, kinematical or dynamical. The left hand side of the commutator becomes
\begin{equation*}\label{eq:ComputationCommutators2}
\begin{split}
[\delta_k,\delta_d]|\Phi_3\rangle&=\delta_k\big(\langle\Phi_1|\langle\Phi_2|D_{123}\rangle\big)-\delta_d\big(\hat k_3|\Phi_3\rangle\big)\\
&=\delta_k\big(\langle\Phi_1|\big)\langle\Phi_2|D_{123}\rangle\big)+\langle\Phi_1|\delta_k\big(\langle\Phi_2\big)|D_{123}\rangle\big)-\hat k_3\big(\langle\Phi_1|\langle\Phi_2|D_{123}\rangle\big)\\
&=-\langle\Phi_1|\langle\Phi_2|(\hat k_1+\hat k_2+\hat k_3)|D_{123}\rangle
\end{split}
\end{equation*}
Using this we have the following cases.

\paragraph{Case $g$ zero or kinematical} The commutator becomes
\begin{equation}\label{eq:ComputationalForm1}
\sum_{r=1}^3\hat k_r|D_{123}\rangle=0
\end{equation}
\paragraph{Case $g$ dynamical} The commutator becomes
\begin{equation}\label{eq:ComputationalForm2}
-\sum_{r=1}^3\hat k_r|D_{123}\rangle=|G_{123}\rangle
\end{equation}
\paragraph{Case two dynamical generators $d$ and $g$} The commutator $[\delta_g,\delta_d]=0$ becomes to cubic order
\begin{equation}\label{eq:ComputationalForm3}
\sum_{r=1}^3\hat d_r|G_{123}\rangle-\sum_{r=1}^3\hat g_r|D_{123}\rangle=0
\end{equation}
These forms of the commutators are convenient for direct calculation.
\subsection{Cubic re-summation formulas}\label{subsec:CubicResummationFormulas}
Based on the cubic momentum re-summation formula \eqref{eq:ResummationFormulaCubic} a number of useful identities follow. 
\begin{subequations}\label{eq:CuReFo}
\begin{equation}\label{eq:CuReFo1}
\sum_r p_r\frac{\partial}{\partial\gamma_r}\mathbb{P}={\scriptstyle\frac{1}{3}}\sum_r p_r{\widetilde p_r}=0
\end{equation}
\begin{equation}\label{eq:CuReFo2}
\sum_r p_r\frac{\partial}{\partial\gamma_r}\mathbb{{\bar P}}=-\sum_r {\bar p}_r\frac{\partial}{\partial\gamma_r}\mathbb{P}={\scriptstyle\frac{1}{3}}\sum_r p_r{\widetilde{\bar p}_r}=-{\scriptstyle\frac{1}{3}}\sum_r {\widetilde p}_r{\bar p}_r
\end{equation}
\begin{equation}\label{eq:CuReFo3}
\sum_r p_r{\widetilde{\bar p}_r}=\mathbb{{\bar P}}\sum_r \frac{p_r}{\gamma_r}-\mathbb{P}\sum_r \frac{{\bar p}_r}{\gamma_r}
\end{equation}
\begin{equation}\label{eq:CuReFo4}
\sum_r h_r=-\frac{\mathbb{P}\mathbb{{\bar P}}}{\Gamma}
\end{equation}
\begin{equation}\label{eq:CuReFo5}
\sum_rh_r\widetilde\gamma_r=-\mathbb{{\bar P}}\sum_r \frac{p_r}{\gamma_r}-\mathbb{P}\sum_r \frac{{\bar p}_r}{\gamma_r}
\end{equation}
\begin{equation}\label{eq:CuReFo6}
\sum_r\widetilde\gamma_r\widetilde\gamma_r\gamma_r=-9\Gamma
\end{equation}
\begin{equation}\label{eq:CuReFo7}
\sum_r\widetilde\gamma_r\widetilde\gamma_r\widetilde\gamma_r\gamma_r=0
\end{equation}
\begin{equation}\label{eq:CuReFo8}
\sum_r \gamma_r\frac{\partial}{\partial\gamma_r}\mathbb{P}=\mathbb{P}\quad\text{ and }\sum_r \gamma_r\frac{\partial}{\partial\gamma_r}\bar{\mathbb{P}}=\bar{\mathbb{P}}
\end{equation}
\end{subequations}
\subsubsection*{A trivial subtlety}
Some care must be exercised when computing commutators with $x$ and $\bar x$ since the transverse momenta are not independent due to conservation. The left and right hand sides of an identity derived using re-summation do not have the same commutator with the transverse coordinates.

\subsection{Hermiticity properties}\label{subsec:HermiticityProperties}
The Lorentz generators of section \ref{subsec:LinearRealization} are chosen mainly for calculational convenience but they do have unusual Hermiticity properties. As it stands, $j$ is anti-hermitian. It can be made hermitian by multiplying with $i$. Then it also becomes Hermitian as a field operator $J$. The generator $j^{+-}$ is Hermitian as a field operator $J^{+-}$. The two generators $j^+$ and $\bar j^+$ are Hermitian conjugates of each other and also as field operators $J^+$ and $\bar J^+$. Lastly, considered as field operators, $J^-$ and $\bar J^-$ are conjugates of each other. 

\section*{Acknowledgment}
I would like to thank Lars Brink and Sung-Soo Kim for discussions at Strings 2011 in Uppsala that inspired me to dig out this old work from the drawers and take a fresh look at it. I also thank Ingemar Bengtsson for reading an early version of the text.
\pagebreak


\end{document}